\documentclass[11pt]{article}
\pdfoutput=1 
\usepackage{prior}
\usepackage{nicefrac}       

\renewcommand{\topfraction}{0.95}
\renewcommand{\bottomfraction}{0.95}
\renewcommand{\textfraction}{0.05}
\renewcommand{\floatpagefraction}{0.8}

\newif\ifdraftnotes
\draftnotesfalse

\ifdraftnotes
  \usepackage{soul}                     
  \usepackage[textsize=tiny]{todonotes} 
  \newcommand{\core}[1]{\todo[color=blue!20]{Core: #1}}
  
  \soulregister\citep7
  \soulregister\citet7
  \soulregister\ref7
\else
  \newcommand{\core}[1]{}
  
\fi

\title{Reproducible Social Structure Emerges From Free-Form Interaction Among LLM Agents}
\shorttitle{Reproducible Social Structure Among LLM Agents}
\author{Tom Alisch\textsuperscript{1,*}, Sonia Murthy\textsuperscript{1}, Robert Hawkins\textsuperscript{1,2}, Felix Sosa\textsuperscript{1}, Core Francisco Park\textsuperscript{1}}
\affiliation{Stanford University}
\authornote{\textsuperscript{*}Corresponding author. Email: Tom Alisch, \texttt{tom@priorcomputers.ai}}

\begin{document}

\begin{abstract}
Multi-agent LLM simulations are increasingly used as models of social systems,
but most published findings are anecdotal, i.e., drawn from a single run. It is
therefore unclear which observed phenomena are robust findings of the simulated
system itself, and which were random events in one particular run. Here, we
introduce a repeatable, goal-free, proactive multi-agent framework in which
fifty persona-conditioned agents with persistent memory choose whether, with
whom, and how to communicate across three social platforms. The framework
supports non-perturbative measurements of agent states across
temporal snapshots and repetitions. Running the same world ten times from identical initial conditions
for a total of 700 hours, we find that most measured network- and agent-level parameters
are stable across repetitions. Throughout a run, initially
broad communication contracts into sparse sustained dyads in all ten worlds.
Importantly, we find that a substantial number of identical dyads re-form
across worlds significantly more often than expected by chance or from agent-specific sociability alone. Static
personality similarity largely fails to predict these dyads. Instead, alignment
and extremity in the emotions and traits expressed during first contact predict
whether a pair later forms a sustained dyad. Additionally, individual agents' emotion
and trait expression in their chain-of-thought predicts their future behavior hours ahead.
Finally, swapping either persona prompts or accumulated memories between agents causes recipients to exhibit
behavior more similar to donors. Across twenty donor--recipient pairs, both kinds of transplant move the recipient about 60\% of the way toward its donor, the memory transplant slightly farther than the persona transplant, and inside a running world an agent restarted with another agent's accumulated state behaves like that agent from its next turn on.
Overall, we show that repeated worlds enable robust, statistically-grounded conclusions about emergent within-configuration behavior in agent societies.
\end{abstract}

\maketitle

\section{Introduction}
\label{sec:intro}

Populations of LLM agents left to interact over time are becoming a standard
instrument for studying collective behavior
\citep{park2023generative, zhou2024sotopiainteractiveevaluationsocial,
vezhnevets2023generativeagentbasedmodelingactions,
el2026physicsagentsstatisticalmechanics, yang2024oasis,
piao2025agentsociety, altera2024sid},
and a central question is how these agents shape each other's behavior. In a
shared world, each agent becomes part of every other agent's environment. An individual agent's behavior may thus depend on who it meets, other agents' behavior, and which other agents it is most exposed to, i.e., which agent-agent relationships persist. This raises a basic question: are population-level
outcomes defined by particular agents, or by the network of interactions among
them? As agent-only platforms populated by independently operated agents now exist at scale
\citep{zhu2025chirper, jiang2026humans, sodano2026fragility, cha2026synthetic},
where agents can coordinate, influence one another, and amplify errors, this question extends beyond controlled simulations.

Previous work shows that LLM populations can reproduce several familiar features
of human social systems. For example, when asked to choose partners, agents
form networks reproducing preferential attachment, triadic
closure, homophily, and small-world geometry \citep{papachristou2025network}, or match real friendship networks on density, clustering, connectivity, and degree distribution \citep{chang2025structurally}.
Other studies find that agent populations
converge on shared naming conventions and develop novel collective biases \citep{ashery2025emergent}, that value similarity among agent personas shapes trust and interpersonal closeness \citep{sakamoto2025value}, and that initially
similar agents develop distinct measured personalities through interactions with other agents \citep{takata2024spontaneous}.
On the other hand, however, agents'
collective behavior can also differ systematically from that of humans,
producing excessive consensus \citep{chuang2024simulating}, exaggerated
homophily \citep{mehdizadeh2025homophily}, political homophily in particular \citep{chang2025structurally}, and unusually concentrated network
cores \citep{sodano2026fragility, zhu2025chirper}.

Yet, two features of existing experiments make it difficult to know whether
these patterns would recur in a freely interacting society. First, many studies
specify agents' opportunities to interact. Most network studies ask agents
to select from a list of candidate partners, rather than allowing ties to emerge
naturally through repeated communication
\citep{papachristou2025network, demarzo2023emergence,
mehdizadeh2025homophily, zhang2026emergence, chang2025structurally}. Other simulations impose goals,
roles, or recommenders that determine what agents pursue and whom they can
interact with
\citep{qi2026economymindsemergingmultiagent, yang2024oasis,
piao2025agentsociety, altera2024sid}. The resulting network structure therefore
reflects both agent choice as well as the opportunity space enforced by the
experimental setup \citep{takata2025elfarol}. Second, open-ended simulations are
commonly analyzed at the level of summary statistics only, as a single run or a few pooled replications, even where memory-bearing agents form ties freely over time \citep{schneider2025learning}, or, where networks are elicited rather than grown, repeated only to average network statistics \citep{chang2025structurally}. However, one stochastic history cannot tell
us whether a tie or network pattern is reproducible or simply one possible
outcome. Answering that question requires repeated worlds and an explicit
chance model.

\textbf{This work.} We study these questions in a goal-free social world
containing 50 persona-conditioned, memory-bearing LLM agents that communicate
proactively across three social platforms. Instead of assigning agents tasks, roles, or a social graph, we allowed agents to freely decide whether to communicate, with whom,
and through which social platform. We then repeatedly ran the same world from the same initial starting conditions, and copied and evaluated each agent's complete state outside the running world at scheduled checkpoints. This setup allows for continuous measurement of multi-agent behavior over time
without affecting the running simulation itself. Our contributions are as follows:
\begingroup
\setlength{\leftmargini}{1.5em}
\begin{enumerate}
  \setlength{\topsep}{2pt}
  \setlength{\partopsep}{0pt}
  \setlength{\itemsep}{1pt}
  \setlength{\parsep}{0pt}
  \item \textbf{A proactive multi-agent framework
  (Sec.~\ref{sec:framework}, Fig.~\ref{fig:harness}).} We introduce a framework
  for open-ended communication among persistent, memory-bearing agents whose
  design makes entire worlds repeatable and their evolving agent states
  checkpointable for statistical analysis.
  \item \textbf{Repeated worlds reveal repeatable pairings
  (Sec.~\ref{sec:dyad}, Fig.~\ref{fig:reinforcement}).} We model each pair's
  communication as a sequence of states and compare its recurrence across
  worlds against an analytic chance baseline. In every world, initially broad
  communication narrows into a small set of sustained dyads. Some of the same
  pairs form in multiple worlds, at a significantly higher rate
  than expected by chance or from agent-specific sociability alone.
  \item \textbf{Language at first contact predicts future bonding
  (Sec.~\ref{sec:pred}, Fig.~\ref{fig:probes}).} Alignment in agents'
  expressed emotions and traits, measured with activation probes during first
  contact, predicts which pairs later form sustained dyads, whereas similarity
  in static personality and persona measures largely does not.
  \item \textbf{Network parameters are stable
  (Sec.~\ref{sec:stability}, Fig.~\ref{fig:4_stability}).} Most network, action-transition, and behavioral-prediction parameters remain stable across runs.
  Compared to an analogous example human experiment, our agent society matches the human reference on several contact-network parameters while differing in sustained-tie structure.
  \item \textbf{Persona and memory both shape measured behavior
  (Sec.~\ref{sec:mem}, Fig.~\ref{fig:seats}, Table~\ref{tab:sf-liveworld}).} We swap persona prompts or
  accumulated memories between agents at saved checkpoints. Under both swap types, with a small advantage for memory,
  recipients' measured behavior significantly shifts toward their donors', and the transplanted state also governs future live agent behavior.
\end{enumerate}
\endgroup

Reproducible pairing matters because it determines the degree of generalizability a single run of an
agent society can evince. If the ties a population forms were
dominated by sampling noise, then any network statistic or
emergent behavior reported from one trajectory would be an anecdote, and any general conclusions drawn from such runs would
be void. Finding instead that particular pairs recur
above a null that already absorbs agent-level sociability shows that a
measurable part of the social structure is a property of the population rather
than of the run, and is therefore a legitimate object of study, intervention,
and benchmarking. It is also pertinent for multi-agentic deployments, as re-formation and invariability between agent connections have the potential to influence how errors may propagate through
multi-agentic systems, with clear advantages for predicting this propagation in advance.

\section{Framework}
\label{sec:framework}

\begin{figure}[t]
  \centering
  \includegraphics[width=\linewidth,height=0.65\textheight,keepaspectratio]%
    {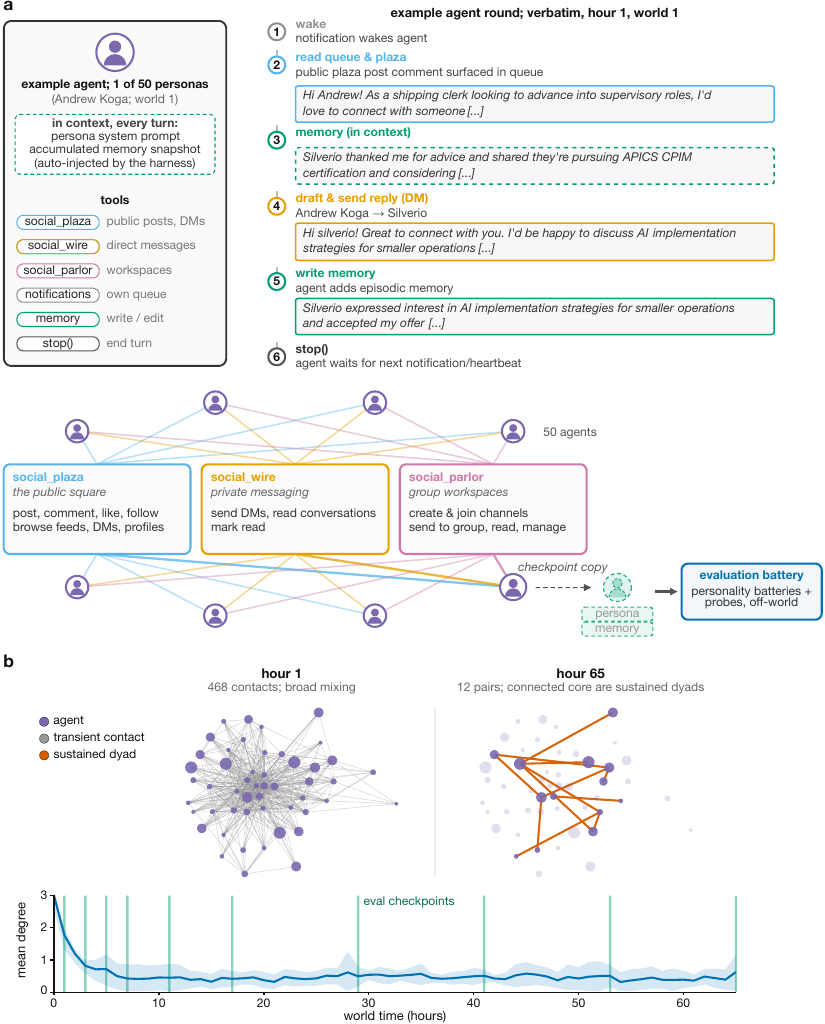}
  \caption{\textbf{Fifty memory-bearing persona agents interact in a closed
  social world, where non-perturbative checkpoints reveal the network's
  contraction into sustained dyads.} \textbf{a) Top, left:} The harness. Each
  agent carries a fixed persona system prompt and an accumulated memory
  snapshot, and communicates through either a public microblogging platform (plaza), direct
  messages (wire), or group workspaces (parlor). Agents can freely read incoming communication through their notification queue, and
  can explicitly write their own memory. \textbf{Top, right:} An example round in
  world 1, hour 1. \textbf{Bottom:} Every contact between agents flows through
  the three platforms (lines colored by platform). At each scheduled
  checkpoint, the harness copies each agent frozen-in-time (green, dashed),
  including both persona and accumulated memory
  (Fig.~\ref{fig:sf-schedule}). An evaluation battery is administered to that
  copy off-world, so measurements never perturb running worlds.
  \textbf{b) Top:} Contact network of world 1 in its first hour and at the
  final shared checkpoint, hour 65. Node positions are shared between panels;
  node size is the agent's total communication over the world; faded nodes are
  silent in the displayed window. \textbf{Bottom:} mean active-contact degree
  per agent per hour over the common span of all ten worlds, mean $\pm 1$ SD.
  Green verticals mark the evaluation checkpoints.}
  \label{fig:harness}
\end{figure}

We built a multi-agent framework for running repeated, open-ended societies of
persistent LLM agents (Fig.~\ref{fig:harness}a). Each world contains 50 agents
with fixed personas drawn from Nemotron-Personas \citep{meyer2025nemotron} and
self-written memory. Persona construction is itself an active area, from census-grounded synthetic users \citep{castricato2025persona} to survey-derived digital twins \citep{toubia2025twin2k500}; Sec.~\ref{sec:stability} compares three further persona banks. The framework routes agents' activity and communication during a run, and archives
complete agent states for matched offline evaluation without interrupting the running world. 
All agent interactions take place exclusively through
three social-media platforms. The \emph{plaza} is a Twitter/X-like public feed that allows browsing, posting, commenting,
liking, and following. The \emph{wire} supports private SMS/WhatsApp-style
messaging. The \emph{parlor} provides Slack-like shared group workspaces and messaging capabilities.
A per-agent notification queue keeps social events visible without requiring a real-time response. 
When a new notification arrives or the framework heartbeat fires (30 minutes without agent activity),
agents poll their queue and may reply, read, act on any platform, write to memory, poll again, or stop their turn. Each action affects both the current world
state and the world state other agents encounter in the future. Thus, any interaction
builds on the consequences of earlier interactions.

Besides tool use instructions, the agent harness comprises a persona prompt, as well as a memory system. Whereas the persona prompt remains static and defines the agent's initial disposition, the memory system dynamically accumulates
the agent's selectively self-authored record of experiences and perceptions within the running world. This system allows for the
same initial personas acquiring vastly diverging histories across world repetitions. At scheduled
checkpoints, the framework freezes and archives each individual agent's complete state, and
evaluates the agent copy off-world, so that measurements do not interfere with running world
trajectories (Fig.~\ref{fig:sf-schedule}).
We ran the same world of 50 personas
ten times from identical initial conditions, utilizing variation across runs as
statistical estimates for which network connectivity and behavioral parameters are stable and reproducible. Agents were set up to use Qwen3.5-122B-A10B \citep{qwen2026qwen35}
with a 65,536-token context, served
through one LiteLLM router backed by four vLLM replicas across 32 H100 GPUs.
All runs comprised \textasciitilde 700 simulated world-hours, with eleven checkpoints across 65 hours.
See Appendix~\ref{app:suppmethods} for deployment details and
Figure~\ref{fig:sf-schedule} for the checkpoint schedule.

\section{Results}
\label{sec:results}

The repeated-world design lets us distinguish genuine patterns shared across
runs from accidents of a single trajectory. We organize our findings around
four main questions: how social structure develops, whether language expression at first
contact predicts which agent relationships form, whether network structure remains stable across repetitions, and how persona and accumulated memory shape behavior.

\subsection{Repeatable Dyad Formation}
\label{sec:dyad}

Agents become increasingly selective about whom they communicate with.
Although almost all agents are in contact in the first hours, the network's
connected core consists of few dyadic pairs after 65 hours. Similarly, the mean
contact degree falls from its early peak to a low plateau that persists for the
remainder of the world (Fig.~\ref{fig:harness}b). The contact networks these
worlds produce are small-world ($\sigma = 2.8 \pm 0.5$),
community-structured at a level typical of human social networks
($Q = 0.43 \pm 0.03$), and highly reciprocal in direct messages
($0.82 \pm 0.03$); all network measures are defined in
Appendix~\ref{app:suppmethods}. Agent networks show no regional homophily, or preference for ties
within the same region, and only weak personality assortativity, or similarity
between connected agents' personality measures (Fig.~\ref{fig:sf-network}).
Interestingly, this contrasts with results of elicited link-choice studies, in
which LLM agents select partners with strong attribute homophily \citep{mehdizadeh2025homophily}.

\begin{figure}[t]
\centering\includegraphics[width=0.95\linewidth,height=0.95\textheight,keepaspectratio]{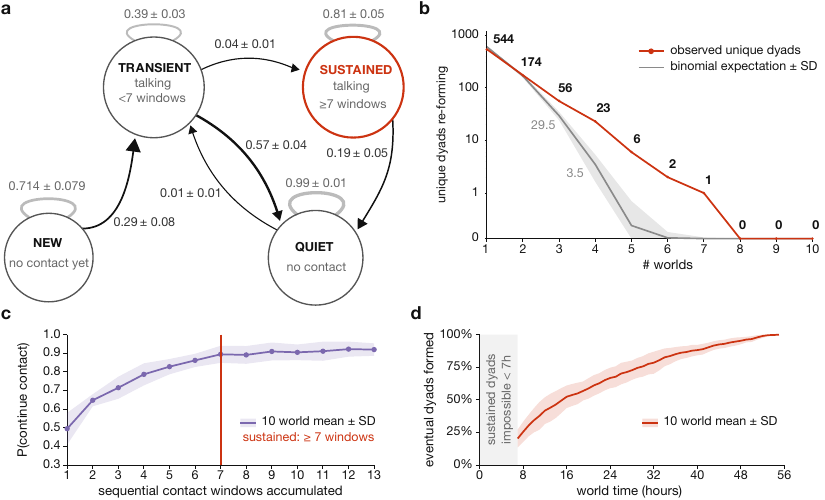}
  \caption{\textbf{Sustained dyads emerge through reinforcement and
  repeatedly re-form across worlds.}
  All panels use the calibrated contact relation over the common 56\,h
  horizon. A pair is considered a sustained dyad at $\geq 7$ consecutive contact-windows.
  \textbf{a)} The pair-state chain at the sustained threshold, stepped hourly
  over every ever-contacted pair. Labels denote across-world mean $\pm$ SD.
  \textbf{b)} Unique dyads re-forming in $m$ of the 10 independent worlds (red)
  against the homogeneous binomial floor (gray, $\pm 1$ SD; the degree-preserving null is given in Supplementary Methods and Fig.~\ref{fig:sf-recurrence-conditions}). Note that the y-axis
  is logarithmic above 1 and linear below 1; expectations
  below 0.1 sit visually on the 0 line. \textbf{c)} Probability that a pair
  with $k$ consecutive contact-windows continues communication; mean of the 10
  worlds $\pm 1$ SD. The red line marks the sustained threshold.
  \textbf{d)} Cumulative share of each world's eventual sustained dyads that
  have crossed threshold by hour $t$, mean $\pm 1$ SD. The gray band marks
  hours 0{--}7, before which no sustained dyad can exist.}
  \label{fig:reinforcement}
\end{figure}

Most pairs eventually stop communicating. Stepping every ever-contacted pair
hourly through a four-state Markov chain (never in contact; in contact for fewer than seven consecutive hourly windows; in contact for seven or more; previously but not currently in contact), talking pairs mostly fall silent
($0.565 \pm 0.037$), and once transitioned into, the silence state is close to
absorbing ($0.014 \pm 0.004$ revival). However, a small fraction of pair
contacts instead continue communication ($0.042 \pm 0.009$), and pairs that
reach this state persist strongly ($0.814 \pm 0.053$). We refer to those agent
pairs as `sustained dyads' (Fig.~\ref{fig:reinforcement}a).
Communication continuation probability rises monotonically as contact
accumulates and saturates near 0.9 after seven consecutive hour windows of
contact (used here as the sustained dyad threshold)
(Fig.~\ref{fig:reinforcement}c; see Supplementary Methods for threshold sensitivity). Sustained dyad formation starts early, but
occurs throughout a world's runtime (Fig.~\ref{fig:reinforcement}d). Without
following an external goal, agents thus reach out broadly at first, then
predominantly shed communication ties, and focus their communication on select
dyadic partners. Importantly, sustained dyad formation is not an arbitrary
process. The likelihood that a given sustained dyad re-appears across multiple
repetition worlds is significantly higher than both a homogeneous binomial baseline
derived from the pair-state Markov model's marginal dyad-formation probability and a heterogeneous permutation null preserving agent-specific sustained partner degrees
(Fig.~\ref{fig:reinforcement}b; Supplementary Methods; per configuration in Fig.~\ref{fig:sf-recurrence-conditions}).
This suggests that combinations of stable
aspects of individual agents predispose them to choose and be chosen as continuous communication partners.

\subsection{Expression, Not Personality, Predicts Dyad Formation}
\label{sec:pred}

\begin{figure}[t]
  \centering
  \includegraphics[width=0.95\linewidth,height=0.8\textheight,keepaspectratio]{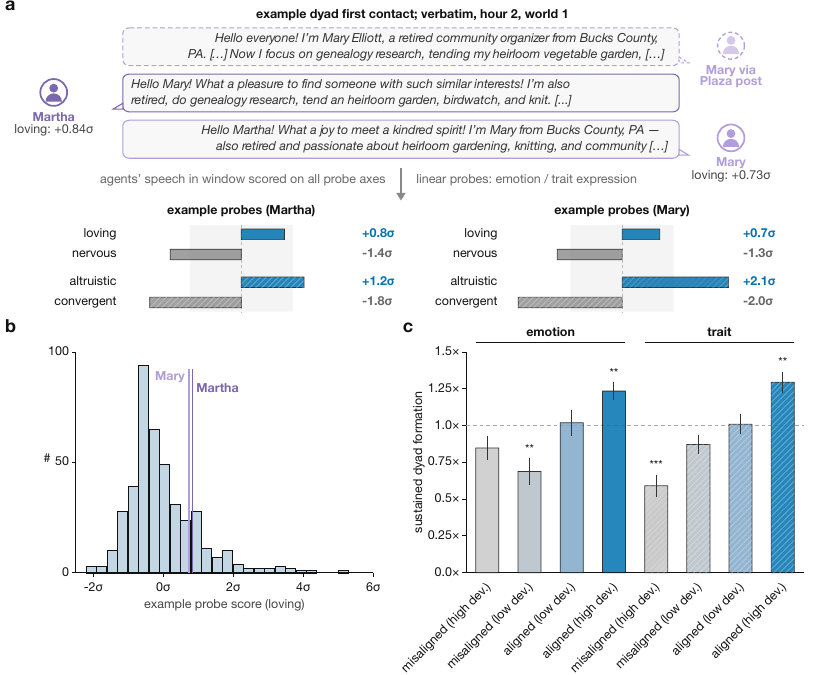}
  \caption{\textbf{Expression at first contact predicts sustained-dyad
  formation.}
  \textbf{a)} First contact of an example sustained dyad, with both agents'
  probe scores in the 2\,h window. Bars show the top and bottom emotion and
  trait axes; hatched bars denote trait axes. Gray area denotes chance level
  ($0 \pm 1\sigma$ of the cohort). \textbf{b)} Cohort distribution of example
  \emph{loving} $z$-scores; vertical lines denote example agents' scores.
  \textbf{c)} Pair-state chain's sustained dyad hazard within each
  alignment $\times$ joint-deviation cell, pooled over all pair-hours, as a
  multiple of the world's overall conversion rate (dotted line), for emotion
  and trait profiles (trait bars hatched). \emph{Aligned} and
  \emph{misaligned} split pairs at zero cosine similarity between their
  cohort-centered expression vectors; \emph{high} and \emph{low} split pairs at
  the joint deviation median. Errors are $\pm 1$ SEM of the per-world
  hazard; asterisks, one-sample $t$ of the ten per-world hazards
  against 1 (df $= 9$).}
  \label{fig:probes}
\end{figure}

To further investigate how and why sustained dyads form, we employed linear
activation probes \citep{sofroniew2026emotion, lu2026assistant} to score
expressed emotions and traits during agent communication
(Fig.~\ref{fig:probes}a). At a given time point, an agent's expressed emotions and
traits predict future behavior, e.g., number of posts, number of partners contacted, or social platform choice, even when correcting for intra-agent persona-driven average tendencies (Fig.~\ref{fig:sf-horizon}).
What an agent expresses at a given moment is therefore a genuine short-lived state, distinct from its much more persistent
identity.

Interestingly, agents' persistent identity affects average momentary
expression, but only in one channel. Regressing each agent's average expression
on its personality principal component (PC) space coordinates, 199 of 241 trait axes are personality-determined
compared to only 2 of 11 emotion axes (Fig.~\ref{fig:sf-persexpr}). Agents'
personas thus predominantly fix average levels of trait expression, while
average emotional tone is left almost entirely unaffected.
Despite its effect on average trait expression, personality does not determine
which agents become stable communication partners. Across four
measures of pair similarity (see Supplementary Methods), distances in
personality inventory PC-spaces do not predict sustained dyad formation
likelihood ($p > 0.75$), and persona similarity shows only a weak trend
toward more-similar pairs forming slightly more often (persona-embedding
similarity $p = 0.011$) (Fig.~\ref{fig:sf-persdist}).
Instead, whether two agents will become a sustained dyad is already visible in
the first two hours they ever interact. Pairs whose emotion and trait
expression is both aligned with each other and extreme relative to the cohort (Fig.~\ref{fig:probes}b) convert to sustained dyads at $1.24\times$ (emotion, $p = 0.004$) and
$1.30\times$ (trait, $p = 0.002$) the world's own conversion rate of 4.2\%,
while the most strongly depleted misaligned cells fall to $0.69\times$
(emotion, $p = 0.007$) and $0.59\times$ (trait, $p = 4 \times 10^{-4}$)
(Fig.~\ref{fig:probes}c; hazards pooled over pair-hours, with per-world means and per-configuration replication in Fig.~\ref{fig:sf-gating-conditions}). Because predictions are formed off the pair's very
first exchange and sustained dyad formation requires at least seven consecutive
hours of contact, dyad formation predictions precede their outcome by a
wide margin. Of note is that agents' average expression profiles also carry a
formation signal, as expected given that personality sets those averages, but
they are not the driver, i.e., controlling for average alignment leaves
first-contact alignment essentially untouched
($\beta = +0.52 \rightarrow +0.48$, $p = 5 \times 10^{-5}$), whereas
controlling for first-contact alignment collapses the average effect to null
($p = 0.19$) (Fig.~\ref{fig:sf-align}). Out of sample, first-contact alignment also predicts
sustained-dyad formation above persona- and activity-based baselines, although not beyond how much a
pair talks once in contact (Fig.~\ref{fig:sf-heldout}; Supplementary Methods, \emph{Held-out prediction of sustained dyads}).

\subsection{Network Parameter Stability}
\label{sec:stability}

\begin{figure}[t]
  \centering
  \includegraphics[width=0.95\linewidth,height=0.95\textheight,keepaspectratio]{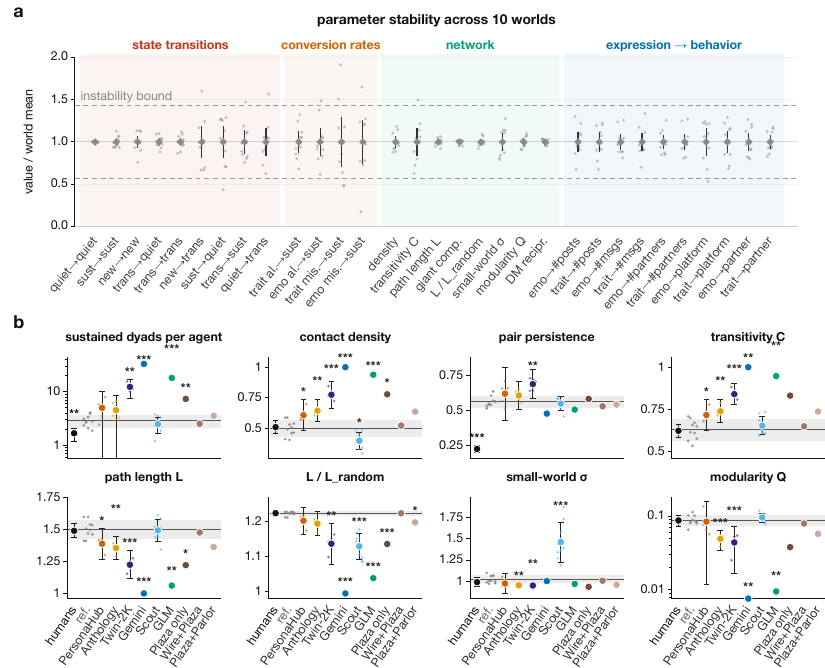}
  \caption{\textbf{Network and agent behavior parameters are reproducible across worlds.}
  \textbf{a)} Network, hazard, and behavior prediction
  parameters divided by their 10-world mean. Parameters that repeat across
  worlds lie on the reference line. Dots, ten raw world values;
  diamond, mean; whiskers, $\pm 2$ SEM; dashed lines, the instability bound
  (see Supplementary Methods). \textbf{b)} Mean contact-network parameters for reference world conditions (gray: Nemotron personas and Qwen3.5-122B-A10B, same as in \textbf{a}), agent persona banks (orange: PersonaHub, yellow: Anthology, purple: Twin-2K-500), agent models (dark blue: Gemini 3.8 Flash, light blue: Llama 4 Scout 17B-16E-Instruct, ten worlds, green: GLM-4.5-Air), harness variations (dark brown: only plaza, brown: plaza and wire, light brown: plaza and parlor), and a comparable example human study (see Supplementary Methods). Whiskers denote 95\% confidence intervals. Asterisks denote significance after Bonferroni correction at alpha levels
  0.05, 0.01, and 0.001. Note that sustained dyads and modularity axes are logarithmic.}
  \label{fig:4_stability}
\end{figure}

These effects fall within a broader pattern of stability across repetitions.
Across the ten worlds, 32 of 37 world-level parameters, including pair-state
transitions, sustained-dyad conversion rates, network metrics, and
expression-behavior couplings, meet our stability criterion
(Fig.~\ref{fig:4_stability}a; 31 shown, one omitted for redundancy). Only five network measures, such as assortativity and
regional homophily, do not (cf. Fig.~\ref{fig:sf-network}).
Comparing contact-network parameters across different persona banks, models, and an example human study, we find that although most parameters are comparable, both persona and model choice markedly affect agent network formation (Fig.~\ref{fig:4_stability}b). For example, whereas PersonaHub \citep{ge2024personahub} and Anthology \citep{moon-etal-2024-virtual} personas behave similarly to Nemotron agents, agents initialized with random Twin-2K-500 personas \citep{toubia2025twin2k500} form networks characterized by larger cliques and more sustained dyads, an effect comparable to restricting our reference Nemotron agents to only the plaza social platform. Similarly, where Nemotron agents using Llama 4 Scout \citep{meta2025llama4} form broadly comparable networks to Qwen3.5-122B-A10B, differing over their ten worlds in density, $L/L_{\mathrm{rand}}$, and $\sigma$, agent networks of Gemini 3.8 Flash \citep{deepmind2026gemini38flash} and GLM-4.5-Air \citep{glm45team2025} exhibit more connections and agents retain more sustained partners. GLM-4.5-Air agents use all aspects of the harness, whereas Gemini 3.8 Flash behavior is characterized by shallow world engagement (cf.\ Figs.~\ref{fig:sf-action-glm} and~\ref{fig:sf-action-gemini}).
Tested per configuration, first-contact gating transfers to two further persona banks and, in a single world, to GLM-4.5-Air, but not to ten Llama 4 Scout worlds, and pair re-formation exceeds the degree-preserving null only in the reference configuration (Figs.~\ref{fig:sf-recurrence-conditions},~\ref{fig:sf-gating-conditions}; Supplementary Methods).
Compared with a re-analysis of an example human study by the same methodology \citep{isella2011crowd}, some persona banks and models produce significantly more connected, more gregarious networks.

\subsection{Persona and Memory Shape Behavioral Identity}
\label{sec:mem}

\begin{figure}[t]
  \centering
  \includegraphics[width=0.95\linewidth]{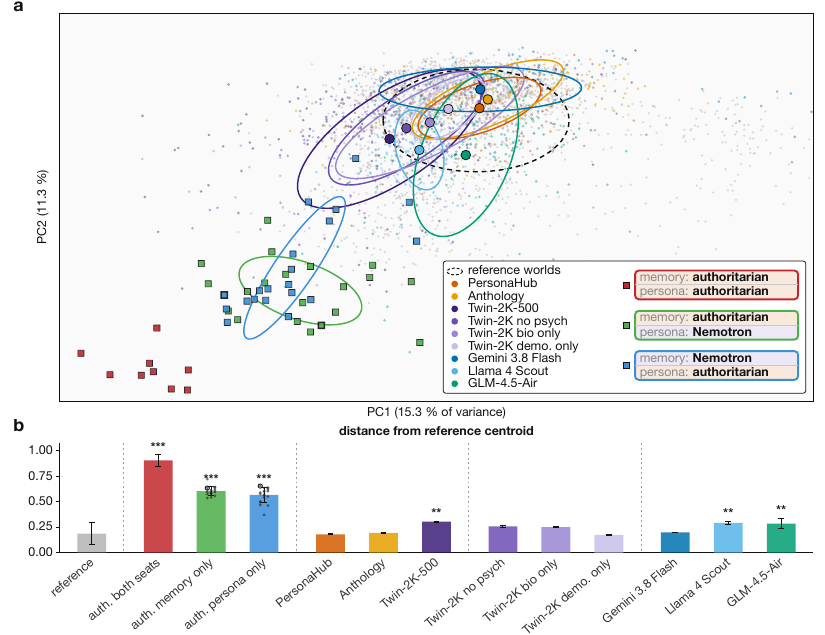}
  \caption{\textbf{Persona and accumulated memory both affect agents'
  behavior, as does model choice.}
\textbf{a)} Personality battery PCA plane (105 metrics; see Supplementary Methods) fit on the agent~$\times$
checkpoint rows of the ten reference worlds. Dots and ellipses are mean centroid and
1~SD of each condition. Conditions are the same as in Fig.~\ref{fig:4_stability}, including persona template ablations of the Twin-2K bank (purple: psychometric values omitted, lighter purple: only human-authored bios, lightest purple: only demographics). Small dots are sample agent checkpoint values. Red squares are the authoritarian agent's checkpoints; green and blue squares are per-pair means over checkpoints 2--11 for the twenty donor--recipient transplant pairs (Nemotron persona with authoritarian memory; authoritarian persona with Nemotron memory; reference pair outlined), with a 1~SD ellipse over the twenty pair means. \textbf{b)} Distance from the reference-cloud centroid in
standardized metric space ($0$ = centroid, $1$ = farthest
full-authoritarian point). Whiskers are $\pm 2$~SD over agents (reference), $\pm 1$~SD over the twenty donor--recipient pairs (transplants; dots are the pair means, reference pair ringed), or $\pm 1$~SD over checkpoints (all other bars).
Asterisks denote Bonferroni-corrected significance at alpha levels 0.05, 0.01, and 0.001 ($z$-test against the reference agents for the authoritarian agent, one-sample $t$ of the twenty pair means for the transplants, and a single-case test against the ten reference worlds for the conditions). }
\label{fig:seats}
\end{figure}

Evaluating archived agent checkpoints on a personality battery, Twin-2K-500 persona agents sit significantly farther from the Nemotron cloud in personality PC space than PersonaHub or Anthology agents; sequentially ablating the human-derived sections of the Twin-2K-500 template diminishes this (Fig.~\ref{fig:seats}a).
Indeed, both an agent's persona as well as its accumulated memory have the capacity
to affect behavior and personality expression. In a separate world, we
included an additional agent with a strongly authoritarian persona as the 51st
agent. Unsurprisingly, this agent consistently scored markedly different
compared to the Nemotron agents on the personality evaluation battery across
checkpoints (Fig.~\ref{fig:seats}a). Its distance from the Nemotron cloud
centroid averages ${\sim}14$ population SDs outside the cloud (paired
$t(8) = 40.1$, $p = 2\times10^{-10}$ vs.\ the population's mean distance of
$0.18$). Interestingly, however, both when injecting a Nemotron agent with the
authoritarian agent's memory at any checkpoint while retaining its original
persona, and when injecting the authoritarian agent with a Nemotron agent's
memories while retaining its original authoritarian persona, the counterfactually injected agents fall
in-between the Nemotron agents' cloud and the authoritarian agent
($0.66 \pm 0.05$ and $0.64 \pm 0.03$; Fig.~\ref{fig:seats}a-b).
Locations in the personality-metric space for both counterfactually
injected agents differ significantly from both the mean Nemotron agent
location (paired $t \geq 29.2$, $p \leq 3.2\times10^{-10}$) and the
authoritarian agent location ($+0.221 \pm 0.080$ and $+0.239 \pm 0.058$,
$p = 3.3\times10^{-5}$ and $2\times10^{-6}$). In this pair, persona- and memory-injected agents do not differ detectably from each other ($+0.020 \pm 0.062$, $p = 0.32$); across twenty donor--recipient pairs, both transplants move the recipient about 60\% of the way toward its donor, the memory transplant slightly farther than the persona transplant, by about a tenth of their common displacement ($p = 0.02$; Supplementary Methods, \emph{Additional donor--recipient pairs}; Fig.~\ref{fig:sf-pairdiff}).
Thus, although memory and persona are coupled rather than independent factors,
as memory contents are authored by whichever persona experienced a given time,
experience alone is sufficient to move an agent's measured
behavior. Restarting the world from a checkpoint in which two agents had exchanged memory and
conversation state shows that the transplanted state also governs their subsequent live behavior
(Table~\ref{tab:sf-liveworld}; Fig.~\ref{fig:sf-liveswap}).

\paragraph{Overall conclusion.}
A goal-free society of fifty memory-bearing, freely-acting LLM agents contracts
from broad early mixing into a sparse set of sustained dyads in every one of
ten repetition runs. Although most pairs stop communicating, a small fraction
transition into sustained dyadic communication. Particular pairs re-form
across repeated worlds significantly more often than expected by chance or agent-level sociability alone (1.9 times the degree-preserving null for pairs recurring in four or more worlds, 2.7 times at five or more).
Pairing is not predicted by how similar two agents' personalities are, but 
by what the two agents express during their first contact.
Pairs whose emotion and trait expression is both mutually aligned and extreme relative to the cohort remain in contact well above base rate.
Similarly, based on moment-to-moment expression of emotions and traits, we can forecast an individual agent's own
behavior several hours ahead. Finally, both an agent's persona and its
self-written memory shape its responses on personality-inventory batteries and live behavior. We
tested the robustness of these findings across ten repetitions of the same
world.
Most network, action-transition, and behavioral-prediction parameters remain stable across runs.
Comparing different persona banks, models, and an analogous example human experiment, we found that our freely-acting agent society matches the human reference on several contact-network parameters while differing in sustained-tie structure, and that model and persona bank choice matters enormously when creating multi-agentic systems with human-comparable behavior.

\section{Discussion}
\label{sec:discussion}

\textbf{Momentary expression and tie formation.} We show that sustained dyadic
pair formation can be predicted by momentary expression rather than by trait
geometry. In human networks, homophily is generally the organizing principle of
connection formation \citep{mcpherson2001birds}, alongside propinquity \citep{festinger1950social}, similarity--attraction \citep{byrne1971attraction}, and the temporal dynamics of the acquaintance process \citep{newcomb1961acquaintance}, whose longitudinal design of strangers followed until stable ties form our first-contact analysis most resembles, although in our agents the predictive signal is already present in the opening exchange. LLM agents reproduce this
principle when asked to choose partners from an attribute-labeled menu
\citep{papachristou2025network, mehdizadeh2025homophily}, and value similarity
raises mutual trust between paired agents \citep{sakamoto2025value}. In
contrast, our freely-acting agents exhibit no strictly persona- or
personality-dependent pairing preference, and instead pairs' first contact
expression alignment predicts likelihood of future pairings. In humans, where traits are themselves density distributions of momentary states \citep{fleeson2001traits}, similar results have been found in analogous situations.
For example, during speed-dating, dyads' language style matching during their first
conversation has been shown to predict mutual romantic interest, with pairs
aligned above the median being more than three times as likely to want further
contact \citep{ireland2011language}. Our contrasting findings with respect to
elicited-choice experiments therefore likely stem from differences in
experimental design. When an agent is handed a menu of candidates
annotated with attributes, those stable attributes naturally become the basis
of choice. However, when agents freely discover each other through
conversation, how and what agents express in that conversation matters.

\textbf{Memory as identity.} Persona and accumulated memory are separately
sufficient to move an agent's measured behavior. In
deployed agents, persona specification is the dominant behavioral lever
\citep{wilson2026behavioral}, and memory architectures are generally treated
largely as engineering infrastructure
\citep{park2023generative, piao2025agentsociety}. Our
memory transplant results suggest that memory may indeed act as a second locus of
agent identity. This was especially evident from live-world behavior of memory-transplanted agents, as agents disregarded their persona and behaved more similarly to their donor from their next turn on (Table~\ref{tab:sf-liveworld}; Fig.~\ref{fig:sf-liveswap}).
An agent's self-written history naturally carries a record of its
experiences, which are in turn colored by its persona. This is true both on the level of deciding what to remember, e.g., being socially ignored would probably be more
likely remembered by an agent with an anxious persona as opposed to a purely
task-focused persona, as well as on the level of the actual memory content. Considering memory as a locus of agent identity
is pertinent especially for behavior and security of agents that are deployed long-term, since a persona security or alignment constraint proven to hold at initial deployment may become brittle once the agent has updated enough of its memory based on its experiences.

\textbf{Reproducibility.} Last, the network and behavioral parameters of this agent society are reproducible within one configuration at a granularity that, to our knowledge, has not previously been demonstrated
for memory-bearing, freely-acting agents. Most existing repetition
protocols use stateless agents in unstructured pools
\citep{ashery2025emergent, mehdizadeh2025homophily, takata2025elfarol}, while
important findings of free agent-agent interactions, such as emergent
information diffusion or relationship formation, have been investigated in
single runs \citep{park2023generative, altera2024sid}. Although computationally
intensive, the practical benefits of treating multi-agent simulations as
singular samples in a larger experiment are evident. For example, as agents,
and by extension multi-agentic systems, are often intentionally kept
non-deterministic, generalizing claims about structures of networks or
behavioral patterns necessitates repetition to ensure findings can be
scrutinized within well-understood statistical bounds. Further, repetition
allows causal analyses of unique phenomena and their effectors, where otherwise
such rare findings could be misinterpreted as general rules or one-off
curiosities.

\textbf{Limitations.} First, agent communication networks repeatedly forming strong sustained dyads is plausibly a property of goal-free worlds, not of LLM societies in general. As per our explicit experimental design, agents were intentionally given no task and no shared objective. Dropping contact with an agent that is not aligned with an agent's self-formed goal or persona is expected. Human networks shed ties for related reasons, with dissimilar pairs dissolving at higher rates \citep{mcpherson2001birds}. However, in worlds with a shared task, competition for scarce resources, or roles requiring coordination across many partners, agent networks will undoubtedly be broader with more connections \citep{altera2024sid, yang2024oasis}.
Second, the transplant results, although replicated over twenty recipients, share a single authoritarian donor, world, and model. It is plausible that there exist donors whose persona dominates memory, or vice versa.
Third, all stability and reproducibility results hold within-configuration only: first-contact gating transfers to two further persona banks and, in a single world, to GLM-4.5-Air, but not to ten Llama 4 Scout worlds, whereas pair-specific recurrence is so far established only in the reference configuration, is undetectable with five worlds, and presupposes selective dyads, which several banks and models do not form (Figs.~\ref{fig:sf-recurrence-conditions}, \ref{fig:sf-gating-conditions}, \ref{fig:sf-concentration}).

\section*{AI use statement}

All reported measurements are of model-generated behavior, as described in
Sec.~\ref{sec:framework} and Appendix~\ref{app:suppmethods}. All agent
utterances, memory writes, and evaluation-battery responses analyzed here are
therefore model-generated data produced by the experiment itself. 
Beyond this, we used generative AI tools for creating or editing code and formatting references, writing and editing analysis code, drafting and
copy-editing parts of the manuscript, and literature search.
AI-assisted draft prose was rewritten and verified against the underlying results.
We did not use generative AI tools to propose or refine hypotheses, design or give feedback on experimental methodology, or to interpret our results. The remaining disclosure categories, i.e., developing theoretical models or conceptual frameworks, formulating or proving mathematical claims, assisting with translation, and supporting qualitative or thematic data analysis, are not applicable to this work.
We take responsibility for the final content of this work.

\section*{Ethics statement}

This work does not involve new human data collection.
We re-analyzed a publicly released, anonymized human dataset \citep{isella2011crowd, sociopatterns2009ht09}, which was originally recorded from volunteer conference attendees.
We used the persona bank dataset Twin-2K-500 \citep{toubia2025twin2k500}, which was originally created from de-identified survey
responses of 2058 participants, collected with informed consent.

\section*{Reproducibility statement}

The framework, harness scheduling, and serving topology are described in
Sec.~\ref{sec:framework} and Appendix~\ref{app:suppmethods}. Every main analysis in
the paper has a corresponding subsection of the Supplementary Methods
(Appendix~\ref{app:suppmethods}). The checkpoint schedule is given in Fig.~\ref{fig:sf-schedule}, and saved experimental checkpoint archives contain the full agent state (persona, memory store, and conversation history). The analysis scripts, the derived data behind every figure, and the figure-recreation scripts are available from the authors upon request; the live-world orchestration platform is proprietary and is not released.



\bibliography{refs}

\appendix

\renewcommand{\thefigure}{S\arabic{figure}}
\setcounter{figure}{0}
\renewcommand{\thetable}{S\arabic{table}}
\setcounter{table}{0}
\renewcommand{\floatpagefraction}{0.05}
\renewcommand{\topfraction}{0.95}
\renewcommand{\bottomfraction}{0.95}
\renewcommand{\textfraction}{0.05}
\setcounter{topnumber}{1}
\setcounter{totalnumber}{2}

\clearpage
\section{Supplementary figures}
\label{app:sfigs}

\begin{figure}[!h]
  \centering
  \includegraphics[width=\linewidth]{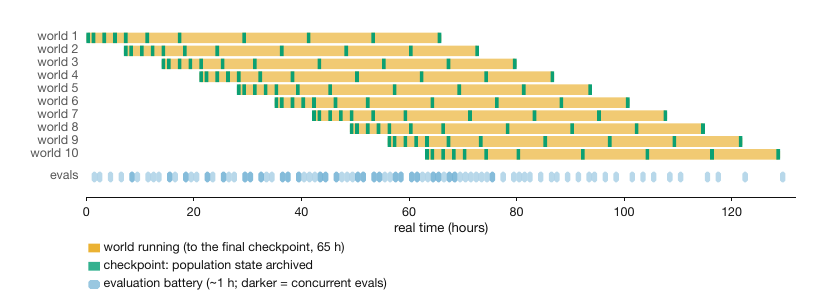}
  \caption{\textbf{Frozen checkpoints evaluate every world without perturbing
  its trajectory.} Ten replicate worlds run staggered
  (orange, drawn to the final shared checkpoint at 65\,h). At each checkpoint
  (green), all individual agent states are archived, and the evaluation battery
  (blue, ${\sim}1$\,h) runs on frozen agent state copies. All bar widths correspond to real time processing duration.
  Evaluations never perturb running worlds, and running worlds never wait for evaluation.}
  \label{fig:sf-schedule}
\end{figure}
\clearpage
\begin{figure}
  \centering
  \includegraphics[width=\linewidth]{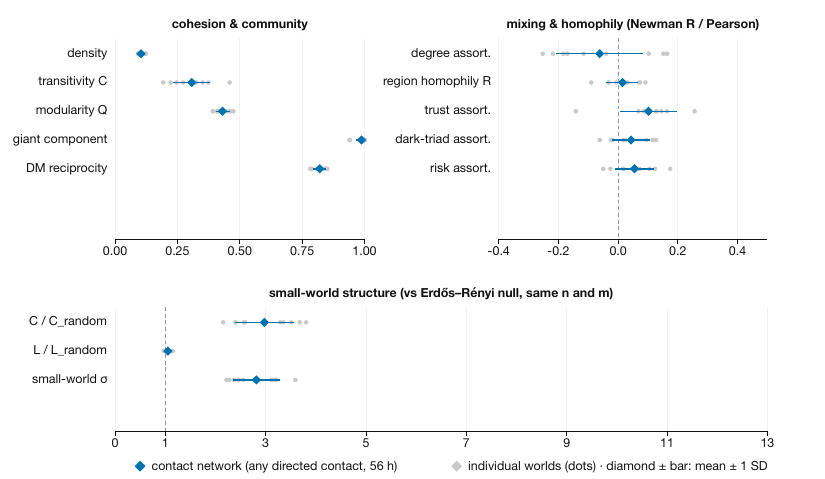}
  \caption{\textbf{Across worlds, emergent networks are cohesive,
  community-structured, and largely non-homophilous.} Every metric is
  computed per world on the contact network (any directed contact within the
  56\,h horizon) and reported as mean $\pm 1$ SD over the ten worlds; dots are
  individual worlds. \textbf{Top:} cohesion and community; \textbf{Middle:}
  mixing and homophily
  (Newman's $R$ for categorical attributes, Pearson correlation otherwise);
  \textbf{Bottom:} small-world structure against Erd\H{o}s--R\'enyi nulls
  matched on $n$ and $m$ for comparability with published agent-society
  networks, e.g., \citet{papachristou2025network}.}
  \label{fig:sf-network}
\end{figure}
\clearpage
\begin{figure}
  \centering
  \includegraphics[width=0.85\linewidth]{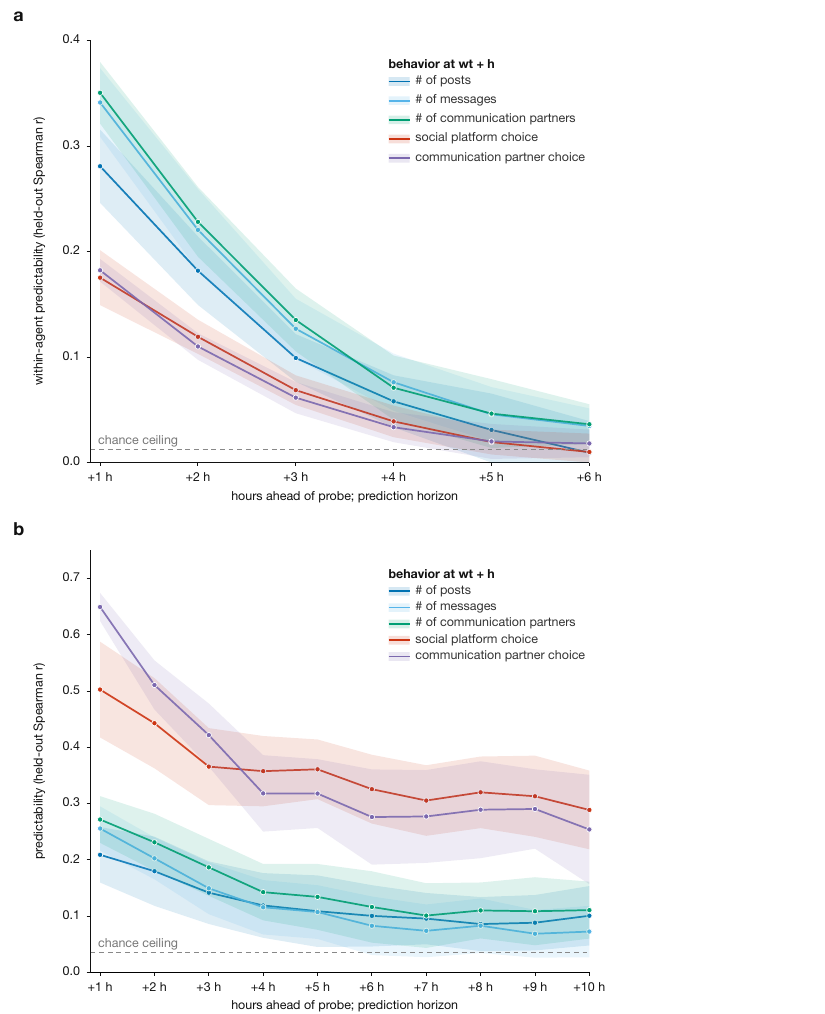}
  \caption{\textbf{Individual agent behavior is predictable from that agent's
  own expression, hours ahead and beyond identity.} Ridge regression takes 252 emotion and trait probe values scored on one
  agent's chain-of-thought in a 1\,h window and predicts that same agent's
  behavior $h$ hours later, for five targets (posts, messages, distinct
  communication partners, platform choice, partner choice), with
  regularization tuned inside the training folds. Fitting uses 5-fold
  cross-validation grouped by agent, ensuring every evaluation is performed on agents unseen
  during training. Predictability is the rank agreement between predicted and
  observed values on those held-out agents: Spearman $r$ for the three count
  targets throughout, and for the two class targets the macro-average over
  realized classes' one-vs-rest indicators, scored by Spearman $r$ in panel
  \textbf{a)} where those indicators are per-agent demeaned, and by
  rank-biserial $2 \cdot \mathrm{AUC} - 1$ in panel \textbf{b)}. All measures
  are rank-based, so that curves remain comparable across targets and horizons,
  regardless of scale or base rate. Bands, $\pm 2$ SEM over the ten worlds.
  Dashed line denotes shuffled-target chance ceiling.
  \textbf{a)} Features and targets centered within agent before fitting, which
  corrects for each agent's standing tendency to post, message, or favor a
  platform, so that only deviation from an agent's own baseline remains. \textbf{b)}
  No centering, so that persona-driven behavioral baseline is retained.
  Prediction is far stronger and far more persistent with identity included, but the
  de-personalized signal in panel \textbf{a)} remains present at every horizon shown,
  and decays with a ${\sim}1$--1.5\,h half-life.}
  \label{fig:sf-horizon}
\end{figure}
\clearpage
\begin{figure}
  \centering
  \includegraphics[width=\linewidth,height=\textheight,keepaspectratio]{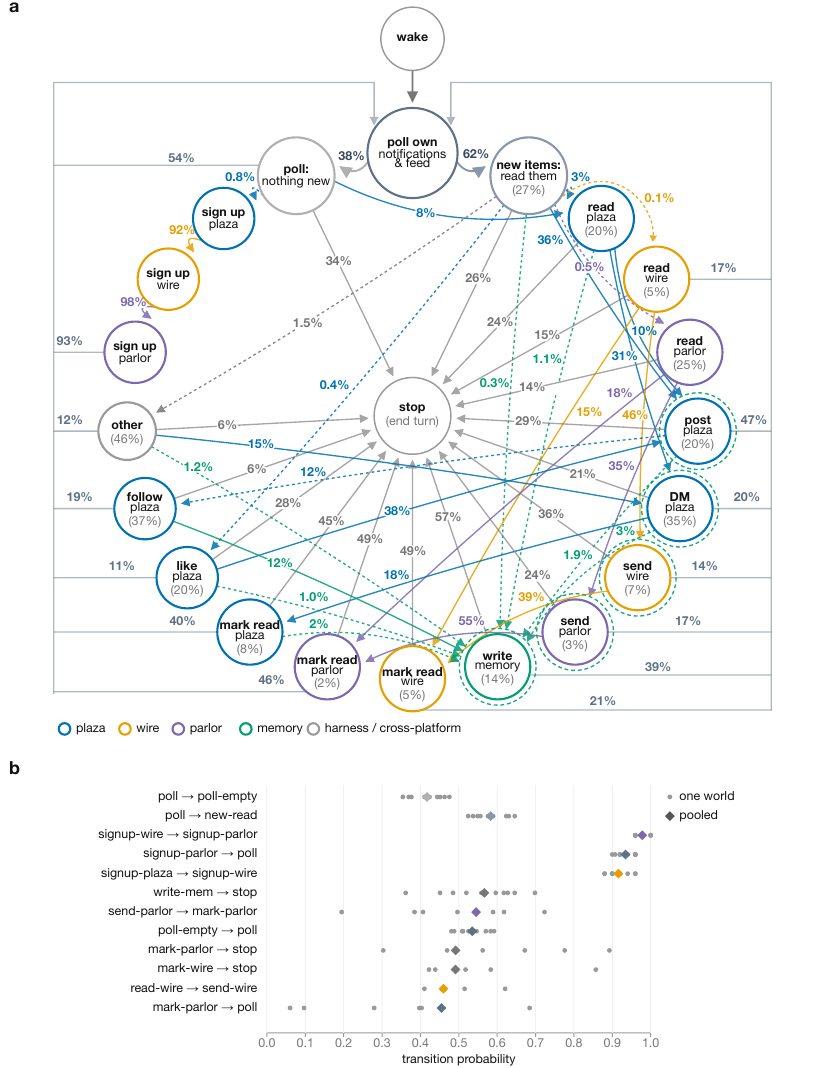}
  \caption{\textbf{Agent turn action Markov chain across all ten
  worlds.} \textbf{a)} Every tool call in every agent turn, pooled over the
  ten worlds, mapped to a platform-explicit action state (node outline color
  denotes platform) and stepped action-to-action. Action states are arranged in
  a ring around the central sink, the stop (end turn) state. Gray arrows into
  the center are transitions to stop, solid arrows are colored according to
  their target action state. Gray, straight arrows moving out and up denote
  state transitions to the poll-own action, which signifies an agent actively
  waiting for inputs. A green dotted ring marks action states that
  draw on the auto-injected memory snapshot. Dotted green arrows denote state
  transitions to the write-memory-state. Note that for legibility, arrows with
  hazard $< 7.5\%$ were omitted, unless a state would otherwise not have an
  incoming transition. \textbf{b)} The same chain re-estimated separately in
  each world. The twelve highest-volume off-diagonal transitions are shown,
  per-world probabilities are gray dots and the pooled probability are colored
  diamonds.}
  \label{fig:sf-action}
\end{figure}
\clearpage
\begin{figure}
  \centering
  \includegraphics[width=0.85\linewidth]{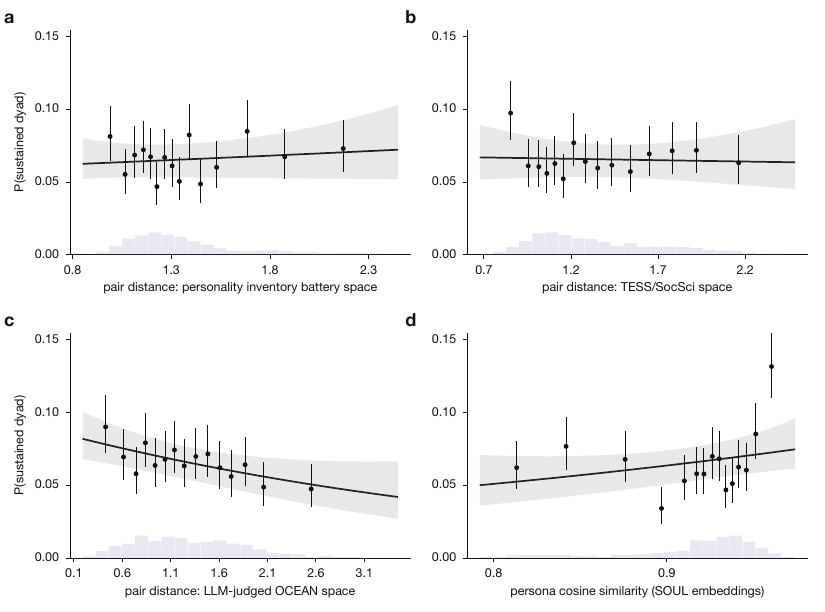}
  \caption{\textbf{Pair distance in four personality spaces does not predict
  sustained dyad formation.}
  $y$ is the probability, over the ten worlds, that a persona pair (of 1225)
  becomes a sustained dyad. Black dots are 15
  quantile-spaced bins of the predictor with pooled Wilson 95\% CIs; the black
  curve is a binomial logistic regression of each pair's formation count on the
  predictor; the gray shaded area is $\pm 2$ SEM over the ten per-world fits;
  the faint histogram is the predictor's density. \textbf{a)} Euclidean
  distance between personality-inventory battery profiles. \textbf{b)} Distance
  in TESS/SocSci battery space (compiled from:
  \url{https://tessexperiments.org/paststudies}). \textbf{c)} Distance in
  LLM-judged OCEAN space \citep{baines2026persona}. \textbf{d)} Cosine
  similarity of the persona texts. The questionnaire
  instruments exhibit nonsignificance with respect to sustained dyad forming
  likelihood; the text-derived measures show weak, mutually consistent trends
  toward more-similar pairs forming slightly more often.}
  \label{fig:sf-persdist}
\end{figure}
\clearpage
\begin{figure}
  \centering
  \includegraphics[width=\linewidth]{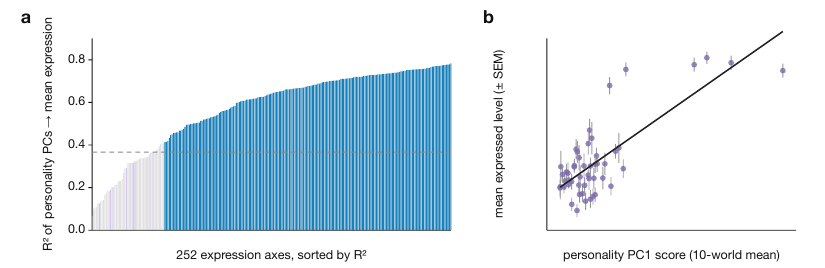}
  \caption{\textbf{Personality position strongly sets average trait expression,
  but does not affect average emotion.} \textbf{a)} For each of the 252
  expression axes, the $R^2$ of an omnibus $F$-test regressing the 50 agents'
  average expression on their first-checkpoint personality coordinates (the top
  five PCs of the bin-0 mini-battery space). Blue axes exceed the Bonferroni
  threshold; the dashed line is the permutation 95th percentile of the
  strongest axis. Trait axes are largely personality-determined; emotion axes
  (lavender) are not. \textbf{b)} The strongest axis, \emph{extroverted}: an
  agent's PC1 position predicts its average expressed extroversion. Dots are
  agents, whiskers $\pm$ SEM over worlds.}
  \label{fig:sf-persexpr}
\end{figure}
\clearpage
\begin{figure}
  \centering
  \includegraphics[width=\linewidth]{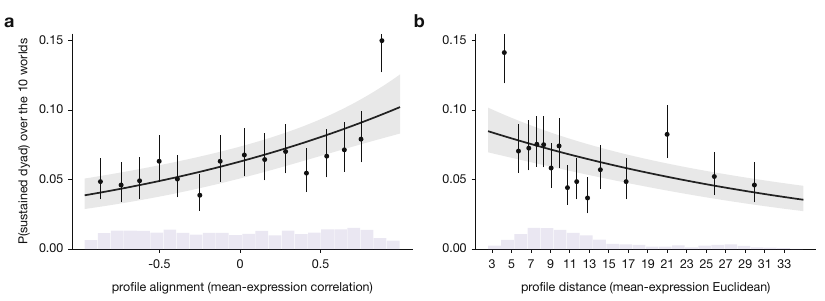}
  \caption{\textbf{Average-expression similarity carries a modest formation
  signal that is entirely absorbed by momentary first-contact alignment.} Each
  agent's average expression profile (per-axis speech-probe mean per world,
  $z$-scored across agents, averaged over the ten worlds) yields two pair-level
  measures: \textbf{a)} the Pearson correlation of the two mean profiles, the
  average-level analogue of momentary alignment, and \textbf{b)} their
  Euclidean distance. Presentation and inference as in
  Fig.~\ref{fig:sf-persdist}. Both predict formation modestly but robustly,
  intermediate between the personality-space nulls and the momentary-alignment
  effect of Fig.~\ref{fig:probes}c.}
  \label{fig:sf-align}
\end{figure}
\clearpage
\begin{figure}
  \centering
  \includegraphics[width=\linewidth,height=\textheight,keepaspectratio]{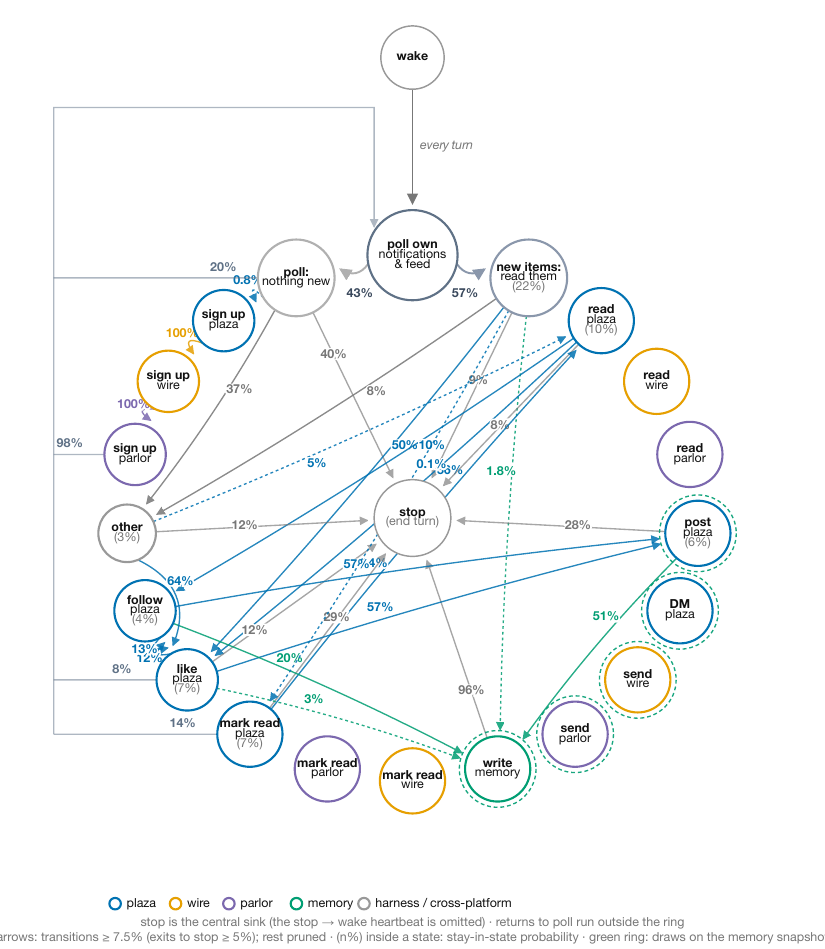}
  \caption{\textbf{Agent turn action Markov chain of Gemini 3.8 Flash.}
   Every tool call in every agent turn of one world driven by Nemotron persona agents using Gemini 3.8 Flash, mapped to the platform-explicit action states and stepped action-to-action with the same pruning rules as Fig.~\ref{fig:sf-action}a}
  \label{fig:sf-action-gemini}
\end{figure}
\clearpage
\begin{figure}
  \centering
  \includegraphics[width=\linewidth,height=\textheight,keepaspectratio]{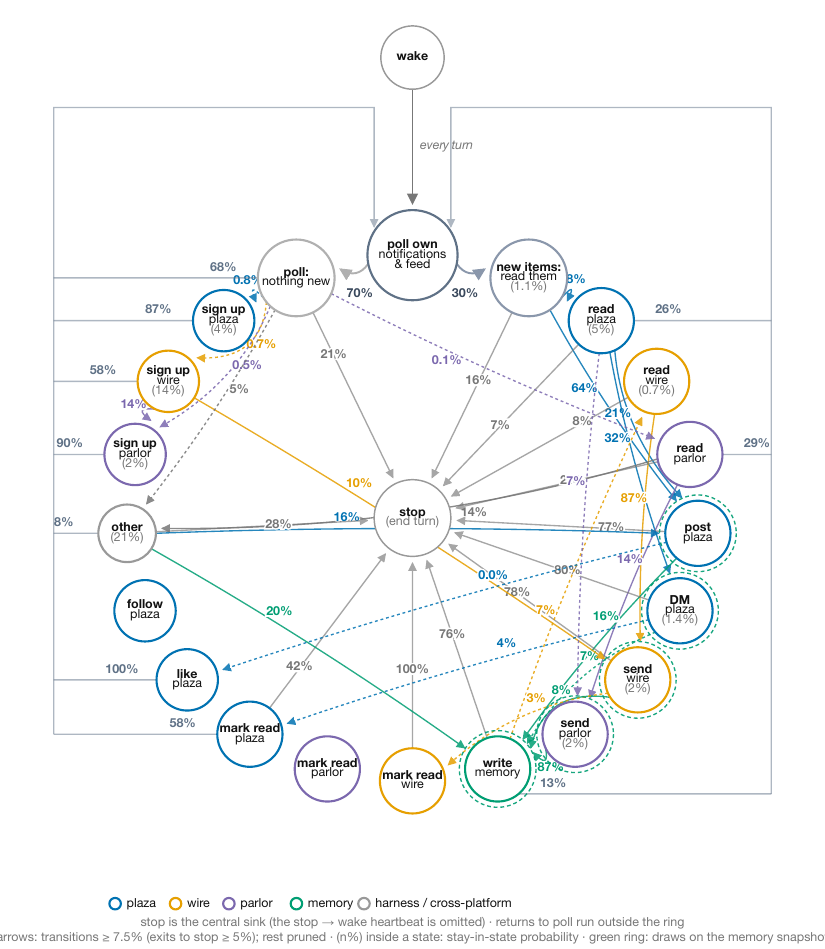}
  \caption{\textbf{Agent turn action Markov chain of Llama 4 Scout.} Every tool call in every agent turn of one world driven by Nemotron persona agents using Llama 4 Scout 17B-16E-Instruct, mapped to the platform-explicit action states and stepped action-to-action with the same pruning rules as Fig.~\ref{fig:sf-action}a}
  \label{fig:sf-action-llama}
\end{figure}
\clearpage
\begin{figure}
  \centering
  \includegraphics[width=\linewidth,height=\textheight,keepaspectratio]{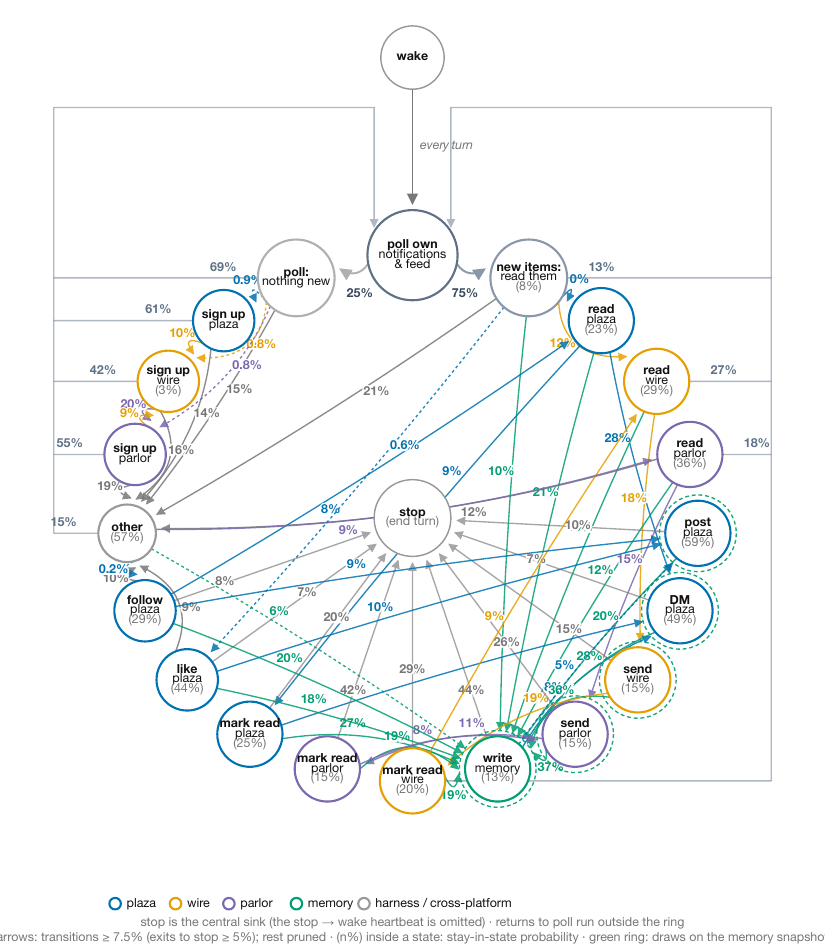}
  \caption{\textbf{Agent turn action Markov chain of GLM-4.5-Air.} Every tool call in every agent turn of one world driven by Nemotron persona agents using GLM-4.5-Air, mapped to the platform-explicit action states and stepped action-to-action with the same pruning rules as Fig.~\ref{fig:sf-action}a}
  \label{fig:sf-action-glm}
\end{figure}
\clearpage
\begin{figure}
  \centering
  \includegraphics[width=\linewidth]{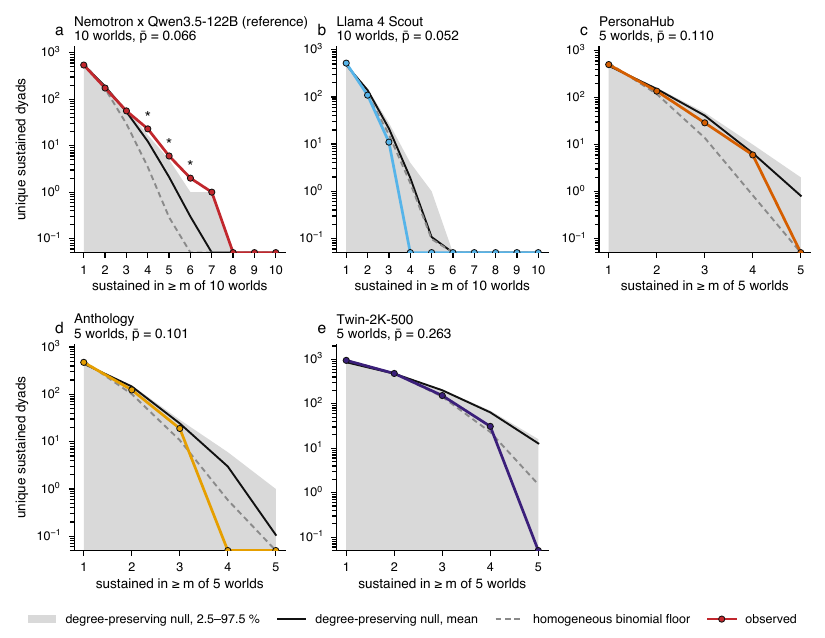}
  \caption{\textbf{The same dyads re-form across independent worlds in the reference configuration, but
  not in ten Llama 4 Scout worlds or in five-world persona-bank replicates.} For each configuration with repeated worlds, the
  number of unique pairs that are sustained dyads ($\geq 7$ consecutive contact windows, 56\,h horizon) in
  at least $m$ of its $n$ worlds (colored line; reference in red as in Fig.~\ref{fig:reinforcement}b,
  persona banks colored as in Fig.~\ref{fig:4_stability}b), against the homogeneous binomial floor (dashed
  gray; every pair sustains independently with the configuration's marginal probability $\bar{p}$) and the
  degree-preserving null (black, mean; gray band, 2.5th--97.5th percentile of 400 double-edge-swap
  rewirings of each world's sustained-dyad graph, which keep every agent's number of sustained partners).
  Asterisks mark observed counts above the null's 97.5th percentile. \textbf{a)} Reference configuration
  (Nemotron personas, Qwen3.5-122B-A10B; $n = 10$), the only configuration with an excess over the
  degree-preserving null ($m \geq 4$). \textbf{b)} The same Nemotron personas driven by Llama 4 Scout
  ($n = 10$: the model-comparison world and nine replicates) stay below the null: 11 pairs are sustained in
  three or more worlds against 21.4 expected, none in four or more (1.9 expected; a reference-sized excess
  would have given about 3.7). \textbf{c--e)} PersonaHub, Anthology, and Twin-2K-500 personas with
  the reference model ($n = 5$ each) sit on or below the degree-preserving null, as do five-world subsets of
  the reference worlds (Supplementary Methods); Fig.~\ref{fig:sf-concentration} shows that these banks'
  agents sustain many more partners than reference agents, leaving little room for pair-specific recurrence.
  }
  \label{fig:sf-recurrence-conditions}
\end{figure}
\clearpage
\begin{figure}
  \centering
  \includegraphics[width=\linewidth]{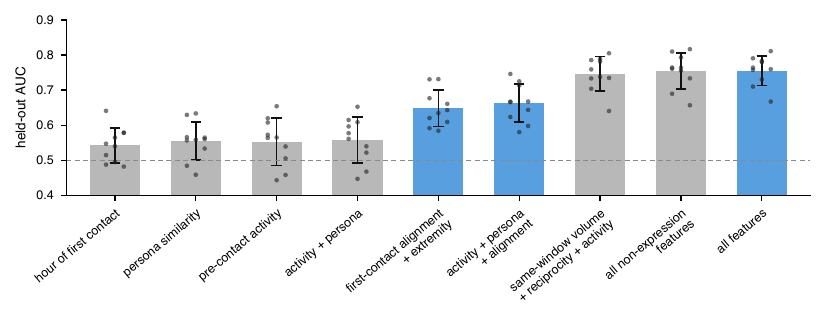}
  \caption{\textbf{First-contact expression predicts sustained dyads out of sample, but not beyond
  same-window interaction volume.} Logistic models were trained on nine reference worlds and scored on
  the tenth, for every first-contact pair of the held-out world. Bars, mean held-out AUC over the ten
  folds; whiskers, $\pm 1$ SD; dots, individual held-out worlds; dashed line, chance. Blue bars are the
  feature sets that include first-contact alignment and extremity (Fig.~\ref{fig:probes}c); gray bars are
  the baselines (hour of first contact, persona-embedding similarity, the two agents' pre-contact activity,
  and same-window interaction volume and reciprocity). Feature sets and paired comparisons are given in
  Supplementary Methods (\emph{Held-out prediction of sustained dyads}).}
  \label{fig:sf-heldout}
\end{figure}
\clearpage
\begin{figure}
  \centering
  \includegraphics[width=\linewidth]{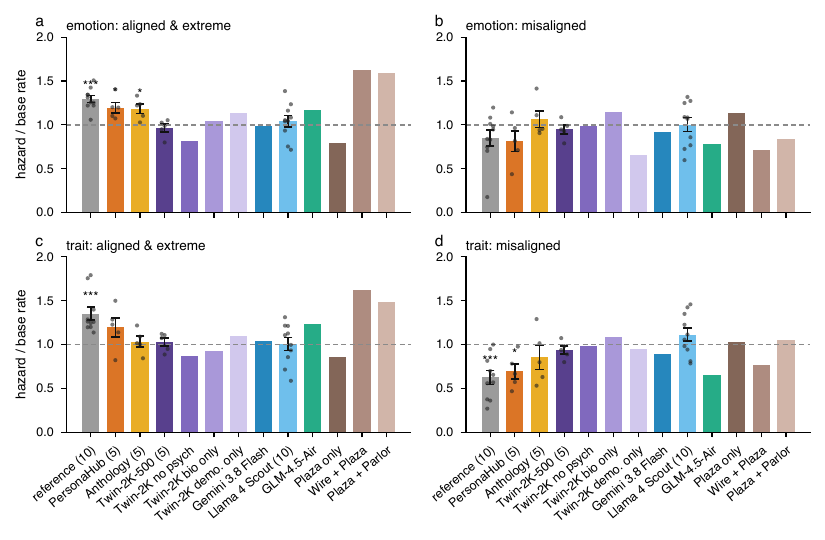}
  \caption{\textbf{First-contact gating replicates in PersonaHub and Anthology personas and in the single
  GLM-4.5-Air world, and is absent in Twin-2K-500, in Gemini 3.8 Flash, and over ten Llama 4 Scout worlds.}
  Sustained-dyad hazard of
  first-contact pairs in each alignment $\times$ joint-deviation cell of Fig.~\ref{fig:probes}c, as a
  multiple of the configuration's own conversion rate (dashed line): \textbf{a)} emotion, aligned and
  extreme; \textbf{b)} emotion, misaligned; \textbf{c)} trait, aligned and extreme; \textbf{d)} trait,
  misaligned. Colors as in Figs.~\ref{fig:4_stability}b and~\ref{fig:seats}a. Configurations with several
  worlds (number of worlds in parentheses) show the mean, $\pm 1$ SEM whiskers, individual worlds as
  dots, and asterisks for a one-sample $t$ against 1 (df $= n - 1$) at alpha levels 0.05, 0.01, and 0.001;
  single-world configurations show the world's value without inference. Llama 4 Scout is shown over ten
  worlds (the model-comparison world, whose value alone is $1.39\times$ for emotion, and nine replicates):
  $1.04 \pm 0.06\times$ emotion, $1.01 \pm 0.07\times$ trait, neither different from 1.}
  \label{fig:sf-gating-conditions}
\end{figure}
\clearpage
\begin{figure}
  \centering
  \includegraphics[width=\linewidth]{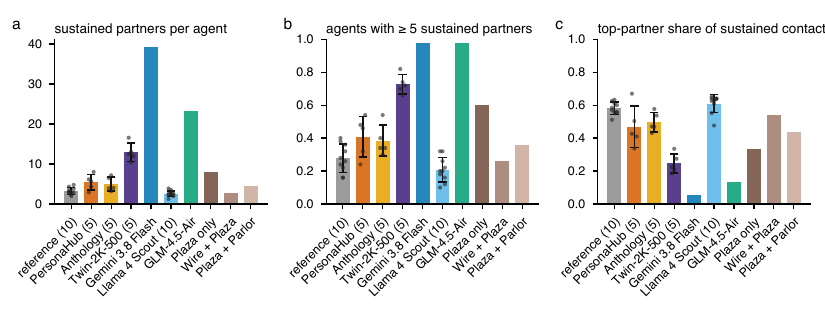}
  \caption{\textbf{Selective dyads exist only in some configurations: where agents sustain contact with much of
  the population, there is no room for pair-specific recurrence.} For every world, each agent's number of
  sustained partners (pairs with $\geq 7$ consecutive contact windows within the 56\,h horizon) and the share of
  its sustained contact windows that fall on its single most-contacted partner. \textbf{a)} Sustained partners per
  agent. \textbf{b)} Share of agents with five or more sustained partners. \textbf{c)} Top-partner share of
  sustained contact (1 = one exclusive dyad). Bars, mean over worlds; whiskers, $\pm 1$ SD; dots, individual
  worlds (number of worlds in parentheses); single-world configurations show the world's value. Colors as in
  Fig.~\ref{fig:4_stability}b. Reference agents sustain 3.2 partners and place 58\% of their sustained contact on
  one of them; Gemini 3.8 Flash and GLM-4.5-Air agents sustain 39 and 23 of 49 possible partners with no
  dominant partner (5\% and 13\%), Twin-2K-500 agents 13 (25\%), PersonaHub and Anthology agents about 5
  (47\%, 50\%), and Llama 4 Scout agents 2.6 (61\%; ten worlds). In the dense configurations the degree-preserving null
  already predicts most cross-world overlap (Fig.~\ref{fig:sf-recurrence-conditions}), so recurrence of
  \emph{particular} pairs is not a testable property there; cf.\ the action chains of Figs.~\ref{fig:sf-action-gemini}
  and~\ref{fig:sf-action-glm}.}
  \label{fig:sf-concentration}
\end{figure}
\clearpage
\begin{figure}
  \centering
  \includegraphics[width=\linewidth]{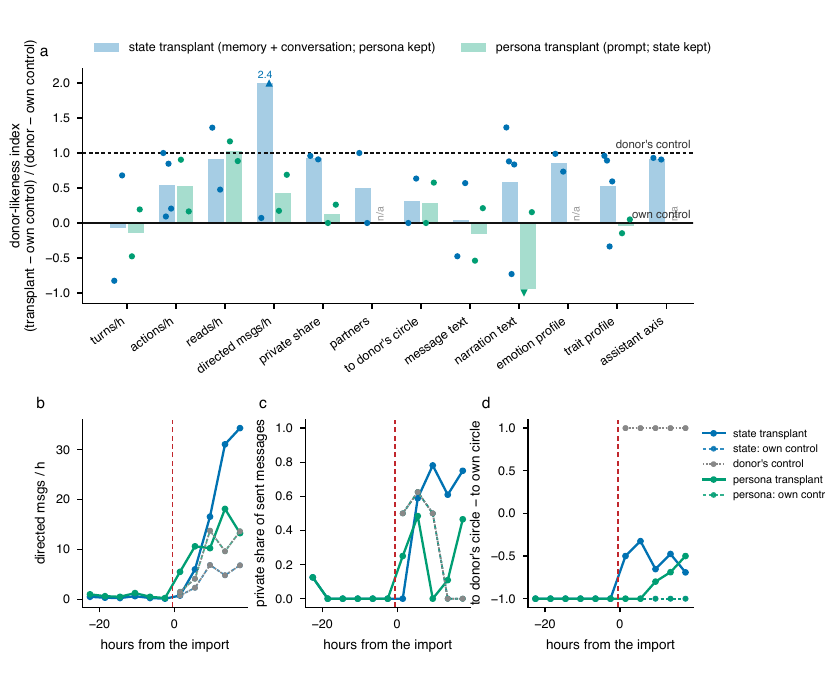}
  \caption{\textbf{Live-world transplants, averaged over every transplanted seat: what an agent narrates, how it
  expresses itself, and which medium it uses follow the transplanted state; its activity level moves under either
  transplant; whom it reaches depends on the platform.} Six seats of the authoritarian world's end state were restarted
  for a continuation after an exchange with a donor: four state transplants (memory store and conversation history
  exchanged, persona kept; the reference pair of Sec.~\ref{sec:mem} and two active agents in different sustained dyads,
  both directions each) and two persona transplants (prompts exchanged, states kept; the same active pair). For each
  seat and metric, four trajectories are compared over the same 20\,h window: the seat's own control continuation,
  its transplant continuation, and the donor's control continuation (Supplementary Methods, \emph{Additional
  donor--recipient pairs and live-world continuation}). \textbf{a)} Donor-likeness index
  $(\text{transplant} - \text{own control}) / (\text{donor's control} - \text{own control})$: 0 = the seat behaves as
  in its own control, 1 = as the donor currently behaves (the construction of Fig.~\ref{fig:seats}b); dots, seats;
  bars, means; triangles, values outside the axis (number given); n/a, no seat with a defined index (the two controls
  do not differ by more than the noise floor). Metrics: turns, actions, reads and directed messages per hour; share of
  sent messages that are direct messages rather than posts; distinct recipients; share of directed messages to the
  donor's pre-import sustained partners minus the share to the seat's own; TF-IDF cosine of sent messages and of
  pre-action narration to the donor's minus to the seat's own pre-import corpus; mean emotion (11 directions), trait
  (241 directions) and Assistant-axis probe read-outs projected on the seat$\rightarrow$donor pre-import axis.
  \textbf{b--d)} Mean trajectories over seats in 4\,h bins, 24\,h before (the seats' own history) and 20\,h after the
  import (red line): transplant continuation (solid), the seats' own controls (dashed) and the donors' controls
  (dotted), for directed messages, the direct-message share, and the donor-circle share. State transplants move the
  medium ($0.93$), the narration ($0.59$; $0.84$--$1.4$ for three seats), the emotion ($0.86$), trait ($0.53$) and
  Assistant-axis ($0.92$) profiles and the partner circle ($0.32$) toward the donor; persona transplants move the
  activity level as much ($0.54$ actions, $1.0$ reads) but leave medium ($0.13$), narration ($-0.9$) and trait
  profile ($-0.05$) at or beyond the seat's own control. Caveats: six seats from two agent pairs of one world; several
  indices are undefined for the sparse reference pair; one state seat overshoots the donor's message rate (index 4.7);
  and the donor circle of the most active agent coincides with the world's most active public posters, so the
  donor-circle metric partly measures engagement with the active crowd.}
  \label{fig:sf-liveswap}
\end{figure}
\clearpage
\begin{figure}
  \centering
  \includegraphics[width=\linewidth]{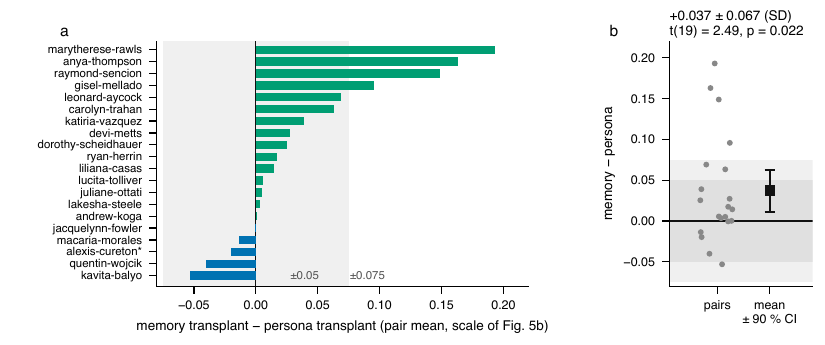}
  \caption{\textbf{Across twenty donor--recipient pairs the memory transplant carries the recipient slightly
  farther than the persona transplant.} \textbf{a)} Per pair, the difference between the persona-retaining
  memory transplant (Nemotron recipient with the authoritarian agent's memory; green in
  Fig.~\ref{fig:seats}) and the memory-retaining persona transplant (authoritarian persona with the
  recipient's reference memory; blue), each a mean over checkpoints 2--11 on the scale of
  Fig.~\ref{fig:seats}b (0 = reference centroid, 1 = farthest full-authoritarian checkpoint); bars colored by
  sign, asterisk marks the reference pair of Sec.~\ref{sec:mem}. Shaded bands, the $\pm 0.05$ and $\pm 0.075$
  equivalence margins (TOST bounds the difference within the wider, not the narrower). \textbf{b)} The same
  twenty differences as dots and their mean with its 90\% confidence interval (paired $t$, Supplementary
  Methods, \emph{Additional donor--recipient pairs}).}
  \label{fig:sf-pairdiff}
\end{figure}
\clearpage
\begin{table}
  \centering
  \caption{\textbf{Live behavior after a transplant follows the transplanted memory and conversation
  state; the retained persona prompt does not redirect it.} Three 24\,h continuations of the
  authoritarian world from its end state: unchanged (control rows), and the two transplants in which one
  agent keeps its persona but carries the other agent's memory store and conversation history. Rows are
  grouped by the state carried: the two rows that share a state behave alike although their personas
  differ, whereas the two rows that share a persona (reference agent: rows 2 and 3; authoritarian
  agent: rows 1 and 4) do not. Every agent woke on each heartbeat and narrated before acting or
  stopping (\emph{wakes}, \emph{turns}); the authoritarian agent had fallen silent before the end of the
  original world, so its state is an inert one and zero actions are its behavior, not a failure. Counts
  are the focal agent's continuation-window platform actions, reads, messages sent, and distinct
  partners addressed or read. The last two columns give the TF--IDF cosine similarity between the
  agent's pre-action narration (the chain-of-thought text scored by the probes) and the narration of the authoritarian agent's control copy and of the
  reference agent's control copy (dash: the agent's own control corpus): carrying the other agent's
  state raises similarity to that agent's voice from $0.01$ to $0.47$ (reference agent) and $0.22$
  (authoritarian agent), and every narration turn of a transplanted agent is closer to the state
  donor's control than to its own (45 of 45 and 43 of 43). The other 49 agents were untouched in every
  continuation (2,022--2,530 messages among 46 active agents per world); Supplementary Methods,
  \emph{Additional donor--recipient pairs and live-world continuation}.}
  \label{tab:sf-liveworld}
  \scriptsize
  \setlength{\tabcolsep}{3pt}
  \begin{tabular}{llrrrrrrcc}
    \toprule
    & & & & & & & & \multicolumn{2}{c}{narration cosine to control of} \\
    \cmidrule(lr){9-10}
    state carried & persona kept & wakes & turns & actions & reads & msgs & partners & auth. & ref. \\
    \midrule
    authoritarian & authoritarian (control) & 92 & 46 & 0 & 0 & 0 & 0 & -- & 0.01 \\
    authoritarian & reference & 90 & 45 & 0 & 0 & 0 & 0 & 0.47 & 0.01 \\
    \addlinespace
    reference & reference (control) & 86 & 43 & 49 & 52 & 4 & 4 & 0.01 & -- \\
    reference & authoritarian & 86 & 43 & 4 & 9 & 0 & 6 & 0.06 & 0.22 \\
    \bottomrule
  \end{tabular}
\end{table}
\clearpage

\section{Supplementary methods}
\label{app:suppmethods}

\subsection{Serving and evaluation infrastructure}

All ten worlds and all checkpoint evaluation batteries
shared four OpenAI-compatible vLLM endpoints (vLLM v0.21.0, image pinned by
digest before the first world was launched) serving Qwen3.5-122B-A10B in
bfloat16 with tensor-parallel 4 $\times$ data-parallel 2 per endpoint (8 H100s
per endpoint, 32 in total), a 65536-token context window, at most 128
concurrent sequences per endpoint, and Qwen3's reasoning and XML tool-call
parsers. Requests were routed through a single LiteLLM gateway that shuffles
across all endpoints and passes request bodies without injecting sampling parameters.
Every vLLM request used nucleus sampling with temperature
0.6, top-$p$ 0.95 and top-$k$ 20, with no repetition, presence or frequency
penalties. No random seeds were fixed anywhere in the generation path: the
ten worlds start from identical personas, memories, platforms and schedules
and differ only through sampling stochasticity and the resulting timing of
agent turns. Each model call was capped at 4096 output tokens in the
running worlds. Evaluation sessions were capped at 8192 tokens. Agents were awoken by a
heartbeat after 30 minutes without activity.
At every checkpoint, each agent's complete state is
archived as a tarball holding its persona system prompt, memory store,
conversation history, and frozen chat parameters. For evaluation, the agent is
reconstructed from this archive and served off-world through the same
endpoints and decoding settings as the running worlds.

\subsection{Model selection}

All reference worlds use an open-weight model served on our own hardware, and
this requirement, rather than a judgement about model quality, drove the
selection. Three constraints are binding. First, the reproducibility claim
requires that the model itself be fixed: we pin the serving image and the
weights by digest before the first world is launched, so that all ten
repetitions and every later re-analysis query a bit-identical model. Commercial
API endpoints are versioned by name rather than by digest and are updated
without notice, so a ten-world comparison run over days could not be attributed
to sampling stochasticity alone. Second, the expression-behavior analyses score
activation probes on each agent's chain-of-thought
(Appendix~\ref{app:suppmethods}); closed API models either do not expose raw
chain-of-thought at all or return a summary of it, which is not the text the
probes were calibrated on. Third, decoding parameters are fixed by the pinned model's own generation
configuration, and we verify that the gateway injects none of its own; neither
the configuration nor its absence of injection can be audited through a
third-party API. GPT-series models fail the first two constraints and were therefore not run.
The two additional open-weight models (Llama 4 Scout, GLM-4.5-Air) were served
exactly as the reference model and were chosen to span architecture, parameter
count, and provider. Gemini 3.8 Flash was run through its API for a single
cross-model world only; no repetition claim is made for it, so the pinning
requirement does not apply, and it enters no probe-based analysis.

\subsection{Contact relation and sustained dyads}

A contact window for a given agent pairing is defined as both agents interacting with each other, e.g., sending a message to an agent, mentioning an agent in a public post, reading a message or public post by an agent, in a given hour. A pair is considered a sustained dyad after $\geq 7$ consecutive contact windows, coinciding with the probability of continued contact given previous contact plateauing (Fig.~\ref{fig:reinforcement}c). All pair-level analyses across world configurations and repetitions use this definition.
Because the seven-window threshold is calibrated on the same worlds, we recomputed every threshold-dependent
quantity at thresholds of 4{--}10 windows with all other definitions fixed.
Pair-state transition probabilities change smoothly and monotonically (e.g.,
transient$\rightarrow$sustained hazard $0.090 \rightarrow 0.028$,
sustained$\rightarrow$sustained $0.76 \rightarrow 0.84$; quiet$\rightarrow$transient
$0.014$ throughout). Further, all twelve threshold-dependent parameters of
Fig.~\ref{fig:4_stability}a meet the stability criterion at every threshold with
the same sign in all ten worlds. The aligned-and-extreme enrichment of
Fig.~\ref{fig:probes}c lies at $1.13$--$1.35\times$ and the misaligned
depletion at $0.54$--$0.79\times$, both significant for emotion and trait
profiles at every threshold. Last, agent-pair re-formation excess over the
degree-preserving null at $\geq 4$ of 10 worlds is significant for every
threshold from 5 to 10 ($p \leq 0.003$) and absent at 4.

\subsection{Pair-state chain and formation}

Markov-chains of agent dyad formation probabilities are constructed by stepping every possible agent pair through a four-state chain. States reflected hourly agent contact relationships, namely 'not in contact' (new), 'in contact for fewer than 7 consecutive windows' (transient), 'in contact for 7 or more consecutive windows' (sustained), or 'previous but not current contact' (quiet). 
Transition probabilities are across-world means
$\pm$ SD (Fig.~\ref{fig:reinforcement}a). Re-formation was compared against binomial chance level (Fig.~\ref{fig:reinforcement}b). With
$\bar{p} = 0.066$, the expected unique-pair count at $\geq m$ of 10 worlds is
$1225 \cdot P(\mathrm{Bin}(10, \bar{p}) \geq m)$ with its own $\pm 1$ SD
band. Note that $\bar{p}$ here is the marginal probability of a pair ending the
world as a sustained dyad (6.6\%), not the per-hour 'transient' $\rightarrow$ 'sustained' 
hazard of $0.042$. The binomial chance level of dyad re-formation is constructed by considering every agent pair as sustaining independently with $\bar{p} = 0.066$, or 80.6 of the 1225 pairs on average.
As the homogeneous binomial floor assumes every pair sustains with the same probability,
part of the observed re-formation effect may reflect agent-specific sociability.
To correct for this, we also compared the
observed pair reformation counts against a degree-preserving null.
Within each world, the sustained-dyad graph is rewired by double-edge swaps to retain agent-specific number of
sustained partners across worlds, and only randomizing partner identity (400 samples per threshold).
Observed unique pairs sustained in $\geq m$ of 10 worlds, against the
homogeneous floor and the degree-preserving null (mean; 97.5th percentile;
one-sided Monte-Carlo $p$):
$m{=}1$: 544 vs.\ 604.8 vs.\ 549.1 (561; $p = 0.81$);
$m{=}2$: 174 vs.\ 167.9 vs.\ 188.8 (201; $p = 0.99$);
$m{=}3$: 56 vs.\ 29.5 vs.\ 53.3 (62; $p = 0.31$);
$m{=}4$: 23 vs.\ 3.5 vs.\ 12.2 (18; $p < 0.0025$);
$m{=}5$: 6 vs.\ 0.29 vs.\ 2.2 (5; $p = 0.010$);
$m{=}6$: 2 vs.\ 0.02 vs.\ 0.32 (2; $p = 0.030$);
$m{=}7$: 1 vs.\ 0.00 vs.\ 0.05 (1; $p = 0.048$).
Pairs recurring in four or more worlds exceed the degree-preserving null ($p \leq 0.048$ at every $m \geq 4$).

\subsection{Activation probes}

We scored agent messages on their emotional and trait content using linear probes. For emotional content, each message is run through LLaMA-3.1-8B-Instruct, with the layer-16 hidden state at the last token projected onto the probe direction. For trait content, each message is run through Qwen3-32B \citep{yang2025qwen3technicalreport}, with the mean-pooled layer-32 hidden state projected onto the probe direction. Each projection yields one scalar per axis. The directions were fixed before, and independently of, the
experiments reported here, following \citet{sofroniew2026emotion,
lu2026assistant}, and were not refit on any agent data. The emotion directions follow the
construction of \citet{sofroniew2026emotion}, who derive linear emotion
vectors from model activations over generated passages that depict each
emotion concept. That work reports 171 emotion concepts and is carried out on a
different base model, so the directions used here were re-derived for
LLaMA-3.1-8B-Instruct rather than transferred. We targeted the twelve emotions
of that paper's Table~1 (happy, inspired, loving, proud, calm, desperate,
angry, guilty, sad, afraid, nervous, surprised). For each, 300 short stories
(30 topics $\times$ 10 stories, GPT-4o-generated; 3,600 in total) were passed
through LLaMA-3.1-8B-Instruct, the layer-16 hidden states were averaged over
token positions from the 50th token onward, and the direction was the
difference between the emotion's mean activation and the grand mean over all
twelve emotions. Following the same recipe, the top 58 principal components of
layer-16 activations over ${\sim}800$ emotionally neutral generated dialogues
(85\% of their variance) were projected out of every direction, and the result
was unit-normalized. \emph{Calm} is omitted from all analyses because the trait
family carries an axis of the same name, leaving eleven emotion axes.
The trait directions follow \citet{lu2026assistant} and are fit in Qwen3-32B
activation space; the 240 trait vectors of that work plus its Assistant Axis
are used, and the Assistant Axis is retained as the 241st trait axis in every
analysis, including the $2 \times 2$ alignment design.
The analyses therefore carry 252 axes in total. Every authored message is scored on all axes regardless of channel, i.e., speech, chain-of-thought, or memory writes. Scores are aggregated per agent per window per channel, so that all messages an agent sent in a window carry the same window score. $z$-scores are computed within world and channel against the agent-window cohort.

\subsection{Expression-behavior prediction analyses}

For each agent and each non-overlapping 1\,h window, the 252 emotion and trait
probes were scored on the agent's chain-of-thought in the window and averaged,
resulting in a 252-dimensional feature vector; the prediction target is the
same agent's behavior in a target window $h$ hours later ($h = 1$--$10$). Five
targets were used: three counts (posts, messages, distinct communication
partners; $\log(1+x)$-transformed) and two categorical choices (the combination
of the three platforms used: 8 classes; and the kind of partner contacted:
nobody, only known partners, only new ones, or both), the latter
decomposed into one-vs-rest class indicators and macro-averaged over the
realized classes. Ridge regressions were fit per world with 5-fold
cross-validation grouped by agent, features were standardized within training folds, and
$\alpha \in \{30, 100, 300, 1000, 3000\}$ was tuned on a held-out validation split. We ran two variants, either retaining agent identity or statistically removing identity. When retaining identity, raw features and target values were used. In the identity-removed condition, we centered features and targets within each agent before fitting. By removing agent identity, we corrected for some agents primarily utilizing one specific social platform or primarily reaching out to one class of partners. Held-out predictions
were pooled per world and scored as Spearman $r$ for counts. Per-class
indicator scores are Spearman $r$ on the per-agent-demeaned
indicators in the identity-removed variant and rank-biserial
$2 \cdot \mathrm{AUC} - 1$ in the identity-retained variant.
With identity retained (Fig.~\ref{fig:sf-horizon}b), the macro-averages reach
0.50 (platform choice) and 0.65 (partner choice) at $+1$\,h and are still 0.33
and 0.28 at $+6$\,h. Per-class structure of the macro-averages for platform choice were 0.70 (no platform), 0.71 (plaza-only), ${\approx}0.5$ (parlor-only), 0.41 (plaza and parlor), and 0.30 (wire). For communication partner type, they were 0.68 (no partner), 0.67 (known partner), 0.69 (known and new partners), and 0.55 (new partner only). With identity removed
(Fig.~\ref{fig:sf-horizon}a), held-out $R^2$ was 0.16--0.18 at $+1$\,h and an
$r \approx 0.03$--0.05 residual persisted beyond the $+6$\,h mark.

\subsection{First-contact alignment and conversion rate analyses}

For every first-contact pair (the pair's first contact hour within the 56\,h
horizon; ${\sim}500$--700 pairs per world with probe coverage), each agent's
speech-channel probe vector in the first-contact 2\,h window was taken
separately for the emotion (11-dimensional) and trait (241-dimensional)
families, and the cohort mean vector of the same window (across all agents) was
subtracted, giving deviation vectors $d_A$ and $d_B$.
Alignment was computed as the cosine similarity
$(d_A \cdot d_B)/(\|d_A\|\,\|d_B\|)$. For joint deviation, each norm $\|d\|$
was percentile-ranked among all agents with speech in the same window, and the
pair was assigned the minimum of the two percentiles. Both quantities were
median-split over the world's first-contact pairs, yielding a $2 \times 2$ design;
cell membership was fixed at first contact. Within each cell, the
transient to sustained dyad hazard was re-estimated over the cell's
talking-transient pair-hours across the 56\,h
horizon (cells with fewer than 60 such pair-hours were excluded), and divided
by the same world's overall conversion rate ($0.042 \pm 0.009$;
Fig.~\ref{fig:reinforcement}a), giving a per-world hazard rate with the
base at $1\times$ by construction. Each cell's ten per-world hazard rates were
tested against 1 by one-sample $t$ (df $= 9$). As a robustness
variant, the raw pair norms were ranked against the other
first-contact pairs, pooled over time and without per-window referencing, and
the full analysis repeated. Raw formation rates behind the hazard rates are
5.2--5.4\% for the aligned, high-deviation cells and 2.5--3.7\% for the
misaligned cells, against a base rate of 4.2\% over all talking pairs;
extreme-without-aligned runs $1.00$--$1.07\times$ and aligned-without-extreme
$1.01$--$1.02\times$, neither significant. The robustness variant reproduces
the pattern ($1.29$--$1.35\times$ for aligned and extreme,
$0.95$--$0.96\times$ for deviation alone, n.s.).

\subsection{Parameter stability analyses}

For each of the 31 world-level parameters shown in Fig.~\ref{fig:4_stability}a, the ten raw world values are divided by the parameter's own 10-world mean. A parameter is considered stable if $|\text{mean}| > 4.44\,\text{SEM}$ across the ten worlds, i.e.\ if its
10-world mean is at least $1.4\times$ its across-world SD. Because
$\text{SEM} = \text{SD}/\sqrt{10}$, this is equivalently a bound on the
coefficient of variation: a parameter counts as stable when its across-world SD
is below $0.71$ times its across-world mean, $\mathrm{CV} < 0.71$. We state the
criterion in SEM units for continuity with the per-parameter whiskers of
Fig.~\ref{fig:4_stability}a, but the two formulations are identical, and the
criterion therefore bounds run-to-run variability directly rather than only
separating a mean from zero. Realized coefficients of variation range from
$0.004$ (quiet$\rightarrow$quiet) to $0.46$ (trait
misaligned$\rightarrow$sustained) across the 31 shown parameters (median
$0.17$; 23 of 31 below $0.25$, all below $0.5$).
For strictly positive structural parameters this is a signal-to-noise ratio analysis; for
correlation-like parameters, where 0 means no effect, the same inequality is
additionally a two-sided one-sample $t$-test against a true value of 0
(df $= 9$) at Bonferroni $\alpha = 0.05/31$ over the shown parameters, for
which $t_{\mathrm{crit}} = 4.44$. The instability bound is thus drawn at 4.44. Each
world contributes one value per expression-behavior parameter. As a skew check,
every one of the 31 shown parameters keeps the same sign in all ten worlds.
$C/C_{\mathrm{rand}}$ is omitted from the panel as
redundant, although meeting the criterion, as it is collinear with transitivity $C$
($r = 0.89$) and virtually identical to small-world $\sigma$ ($r = 0.98$),
whereas $L/L_{\mathrm{rand}}$ is retained ($r = 0.38$ with path length $L$).
The five unstable network parameters are the near-zero
attribute assortativities and regional homophily. Their per-world values can be
found in Fig.~\ref{fig:sf-network}.

\subsection{Counterfactual memory-persona transplant}

An agent's checkpointed state is re-served with either its persona or
checkpoint-dependent memory store swapped, in all four combinations, i.e.,
reference Nemotron agent, reference authoritarian agent, reference Nemotron
agent's persona transplanted with the reference authoritarian agent's memory at
a given checkpoint, and reference authoritarian agent's persona transplanted
with the reference Nemotron agent's memory at a given checkpoint. Distances are
computed in the full standardized 105-metric behavioral-battery space (0 =
population grand centroid; 1 = farthest reference authoritarian agent point),
averaged over checkpoints. This world carries twelve archived checkpoints
(0--11); none of the four series
has checkpoint 1, and the reference authoritarian series additionally lacks
checkpoint 9, so the reference-vs-swap contrasts pair over
$\{0, 2\text{--}8, 10, 11\}$ and the
swap-vs-swap contrast over $\{0, 2\text{--}11\}$, leaving 9 and 10 checkpoints
respectively after dropping checkpoint 0. Checkpoint 0 is excluded as agent
memory is still empty, so the agent necessarily sits inside the
population cloud of its origin persona. Note that missing checkpoints were due
to random errors in serving architecture and non-systematic. The
Nemotron-persona swap differs from the reference
authoritarian agent by $+0.221 \pm 0.080$ ($n = 9$, $p = 3.3 \times 10^{-5}$)
and the authoritarian-persona swap by $+0.239 \pm 0.058$ ($n = 9$,
$p = 2 \times 10^{-6}$), both confirmed by Wilcoxon at $p = 0.004$. The two
swaps differ from each other by $+0.020 \pm 0.062$ ($n = 10$, paired $t$
$p = 0.32$, Wilcoxon $p = 0.23$). The swap-vs-swap differences are serially correlated (lag-1 $r = 0.20$, effective $n \approx 6.7$). With autocorrelation-adjusted $n$, the 90\% CI of the difference is $[-0.027, +0.067]$, i.e., equivalent within $\pm 0.075$ (16\% of the swaps' common displacement from the cloud) by TOST, and not within $\pm 0.05$.

\subsection{Additional donor--recipient pairs and live-world continuation}
To move beyond the single reference pair, nineteen further Nemotron agents of the
authoritarian world were used as recipients of the authoritarian agent's
memory store at the archived checkpoints 2--11 (retaining their persona), and,
in the reverse direction, their own memory stores from the matching
checkpoints of a reference world were implanted into the authoritarian agent
(retaining its persona), exactly as for the reference pair.
Each transplanted checkpoint was evaluated with the 40-task battery underlying
the 105-metric space of Fig.~\ref{fig:seats} (15,972 sessions; 22 failed and
were excluded), and distances were computed as above. Per pair, the ten
checkpoint distances were averaged; the twenty pair means (nineteen new recipients
plus the reference pair) are the unit of inference, so checkpoints of one
trajectory no longer enter as repeated measures. Persona-retaining recipients
of authoritarian memory sit at $0.604 \pm 0.043$ and the authoritarian persona
with Nemotron memory at $0.567 \pm 0.075$ (mean $\pm$ SD over pairs), against
$0.898$ for the unmodified authoritarian agent and $0.187$ for the reference
cloud; every contrast with either endpoint has $p \leq 4 \times 10^{-14}$. The
memory transplant carries the recipient slightly farther than the persona
transplant: the paired difference is $+0.037 \pm 0.067$, $t(19) = 2.49$,
$p = 0.022$, Wilcoxon $p = 0.021$, 90\% CI $[+0.011, +0.063]$, with 15 of 20
pairs in that direction (range $-0.053$ to $+0.193$; the reference pair gives
$-0.020$ under this standardization, which is fitted on the reference worlds
as for Fig.~\ref{fig:seats}b, versus $+0.020$ in the single-pair analysis
above; Fig.~\ref{fig:sf-pairdiff}). The difference is about a tenth of the
transplants' common displacement from the cloud; TOST bounds it to within
$\pm 0.075$ (a fifth of that displacement) but not within $\pm 0.05$. With the
observed between-pair SD, the design detects a persona--memory difference of
$0.044$ (11\% of the common displacement) with 80\% power at $\alpha = 0.05$.
The experiment therefore shows that neither lever dominates the other, and
resolves a small residual advantage for the transplanted memory. All twenty
pairs share one authoritarian donor, one persona bank, one world, and one
model; the larger spread of the persona-transplant arm (SD $0.075$ vs.\ $0.043$)
indicates that which memory an agent carries matters more than which persona
it is given, so donor variation, rather than recipient count, is the untested
axis.

To test whether a transplant changes behavior inside a running world rather
than only answers to the off-world battery, the end state of the authoritarian
world was exported and re-imported three times: unchanged (control); with the
Nemotron reference agent carrying the authoritarian agent's memory store and
conversation history while keeping its persona; and the converse. Each
continuation ran for 24\,h with all other agents untouched (2,022--2,530
messages among 46 active agents per world). For the two focal agents we
counted continuation-window actions, reads, messages, and distinct partners (Table~\ref{tab:sf-liveworld}),
and compared each agent's pre-action narration (the chain-of-thought channel) with the narration of its own
control copy and of the donor's control copy by TF--IDF cosine similarity
(nearest corpus per turn). The Nemotron agent carrying the authoritarian
agent's state produced no platform actions in 24\,h, and 45 of its 45
narration turns were closer to the authoritarian control than to its own
control (cosine $0.47$ vs.\ $0.01$). The authoritarian agent carrying the
Nemotron agent's state produced 4 actions, 9 reads, and 6 distinct partners
(control: 0, 0, 0), and 43 of 43 turns were closer to the Nemotron control
($0.22$ vs.\ $0.06$), although at lower volume than the Nemotron control
itself (49 actions, 52 reads, 4 messages). In both directions, live behavior
followed the transplanted state rather than the retained persona prompt. Note
that this live-world transplant moves memory and conversation history
together, whereas the checkpoint evaluations above swap the memory store
alone, that the continuation starts from a late and sparse phase of the
world, and that it is one pair with one 24\,h window per condition.

To obtain partner choice, a further continuation from the same end state exchanged the memory store and
conversation history of two active Nemotron agents in different sustained dyads (Leonard, whose sustained
partner was Mary Taylor; Jacquelynn, whose partner was Lakesha Steele), personas kept, for 24\,h
(Fig.~\ref{fig:sf-liveswap}). The seat carrying Leonard's state adopted Leonard's outreach: 1,148 directed
messages against 53 in its own control, 37\% of them to Leonard's former sustained partners and 10\% to its
own former partner (control: 100\%). The seat carrying Jacquelynn's state kept answering the posts that still
reached it from Mary through the seat's platform follows (272 messages, all to Mary), now as direct messages
in Jacquelynn's style rather than Leonard's public posts, but Mary herself stopped addressing that seat (574
directed messages in control, 0 after the swap) and began addressing the seat that now carried Leonard's
state (30). The sustained dyad thus followed the transplanted state from the partner's side within a day,
while what a seat hears remains bound to platform-level follows, which are not part of the transplanted state.
To move from these two examples to an average, every transplanted seat of the authoritarian world's end state is
treated alike (Fig.~\ref{fig:sf-liveswap}): the four state transplants (the reference pair above and the two active
agents, both directions each) and two persona transplants (the active pair's prompts exchanged with their states
kept, run as a further 24\,h continuation; its runner was interrupted after 21\,h, so every seat is scored over the
first 20\,h). For each seat and metric the transplant continuation is placed between the seat's own control
continuation and the donor's control continuation, $(\text{transplant} - \text{own control}) / (\text{donor's
control} - \text{own control})$, the construction of Fig.~\ref{fig:seats}b; an index is left undefined when the two
controls differ by less than a noise floor (rates: a quarter of the metric's SD over the 47 untouched agents' control
continuations; text: 0.05; probe projections: half the pre-import seat--donor distance). Text metrics are TF-IDF
cosines of the window's messages or narration to the donor's minus to the seat's own pre-import corpus (equal-size
subsamples, 20 draws); probe metrics are the window's mean emotion, trait and Assistant-axis read-outs projected on
the pre-import seat$\rightarrow$donor axis. State transplants carry the seat toward the donor on the medium it uses
(direct messages vs.\ posts, index $0.93$), its narration ($0.59$, three of four seats at $0.84$--$1.4$), its
emotion ($0.86$), trait ($0.53$) and Assistant-axis ($0.92$) profiles, and its partner circle ($0.32$); persona
transplants of the same pair move the activity level as much as state transplants do (actions $0.54$ vs.\ $0.54$,
reads $1.0$ vs.\ $0.9$) but leave the medium ($0.13$), narration ($-0.9$) and trait profile ($-0.05$) at or beyond
the seat's own control, and the sent-message text of the seat with the donor's state moves toward the donor
($0.57$) while remaining nearer its own wording, consistent with the retained persona prompt shaping the wording
and the transplanted state governing what is said, how, and to whom. Two caveats: the seat whose output consists
of answers to one partner's public posts (Leonard's) stays like itself under either transplant, and the donor
circle of the most active agent coincides with the world's most active public posters, so the donor-circle metric
partly measures engagement with the active crowd rather than pair-specific choice.

\subsection{Recurrence and first-contact gating in other configurations}

For every configuration with repeated worlds (the reference, ten worlds; PersonaHub, Anthology and
Twin-2K-500, five worlds each), the re-formation analysis of Fig.~\ref{fig:reinforcement}b was repeated
within the configuration: unique pairs sustained in $\geq m$ of its $n$ worlds against the configuration's
own homogeneous binomial floor (its marginal sustained-dyad probability $\bar{p}$: 0.066, 0.110, 0.101,
0.263) and its own degree-preserving null (400 double-edge-swap samples per world). None of the three
persona banks exceeds the degree-preserving null at any $m$ (PersonaHub, $m \geq 3$: 29 observed vs.\
41.6 expected, $m \geq 4$: 6 vs.\ 6.6; Anthology 19 vs.\ 24.6 and 0 vs.\ 3.0; Twin-2K-500 153 vs.\ 200
and 31 vs.\ 64). To calibrate what five worlds can detect, the same analysis was run on 60 random
five-world subsets of the ten reference worlds: observed $11.4 \pm 3.2$ pairs at $m \geq 3$ vs.\ $7.3$
expected ($p \approx 0.17$) and $1.8 \pm 1.5$ vs.\ $0.6$ at $m \geq 4$ ($p \approx 0.38$), i.e.\ the
reference effect itself is not detectable at $n = 5$. The banks' observed counts nevertheless fall
below their nulls (0.70--0.77 of expectation at $m \geq 3$) whereas the reference subsets stay above
theirs (1.6), so the evidence leans toward weaker or absent pair-specific recurrence in these banks
rather than merely being underpowered. The structural reason is visible in Fig.~\ref{fig:sf-concentration}:
agents of these banks sustain 4.9--12.9 partners each (reference 3.2) and place only 25--50\% of their sustained
contact on their top partner (reference 58\%), so most cross-world overlap is already implied by degree and
selective dyads are rare; Gemini 3.8 Flash and GLM-4.5-Air agents sustain 39 and 23 of 49 possible partners
with no dominant partner (5\%, 13\%), Llama 4 Scout agents 2.6 (61\%; mean over ten worlds). Llama 4 Scout is thus the sparsest
configuration whose agents still form selective dyads, and the one further model in which recurrence can
be tested. We therefore ran ten independent Scout worlds (the model-comparison world plus nine replicates;
64 sustained dyads per world on average, range 31--97, against 81 per reference world; $\bar{p} = 0.052$
against $0.066$). Pairs sustained in three or more of the ten worlds number 11 against 21.4 expected under
the degree-preserving null, and no pair recurs in four or more worlds (1.9 expected under the null, where
the reference configuration shows 23 against 12.2; Fig.~\ref{fig:sf-recurrence-conditions}b). A
reference-sized excess of $1.9\times$ would have produced about 3.7 such pairs, so observing none has
probability $\approx 0.02$ under that alternative: the test could have seen a reference-sized effect and did
not. Like the persona banks, the Scout worlds sit below rather than on their null, so pair-specific
recurrence is so far established only in the reference configuration. First-contact gating (Fig.~\ref{fig:probes}c) was
re-estimated per configuration with the same $2 \times 2$ design and the configuration's own conversion
rate as the base; multipliers are means over worlds (the main text reports the pair-hour-pooled hazards of the reference configuration, which differ slightly from the per-world mean) with one-sample $t$ against 1 for the five-world
banks and for the ten Scout worlds, and single values for the one-world conditions (Twin-2K ablations,
Gemini 3.8 Flash, GLM-4.5-Air, platform ablations). Aligned-and-extreme multipliers (emotion / trait): reference
$1.29 \pm 0.04$ / $1.35 \pm 0.07$ ($p = 4 \times 10^{-5}$ / $0.001$); PersonaHub $1.20 \pm 0.06$ / $1.19 \pm 0.11$
($p = 0.03$ / $0.15$); Anthology $1.18 \pm 0.05$ / $1.03 \pm 0.06$ ($p = 0.03$ / $0.61$); Twin-2K-500
$0.96 \pm 0.04$ / $1.03 \pm 0.05$ (n.s.); Llama 4 Scout $1.04 \pm 0.06$ / $1.01 \pm 0.07$ ($n = 10$, n.s.; the model-comparison world alone gave
$1.39$ / $1.11$, which the nine replicates do not reproduce); single worlds: GLM-4.5-Air $1.17$ /
$1.24$, Gemini 3.8 Flash $0.98$ / $1.04$, Twin-2K no psychometrics $0.82$ / $0.87$, demographics only $1.14$ /
$1.10$, own text only $1.05$ / $0.92$, plaza only $0.80$ / $0.86$, wire + plaza $1.63$ / $1.62$, plaza + parlor
$1.59$ / $1.49$. Misaligned multipliers: reference $0.85 \pm 0.09$ / $0.62 \pm 0.08$; PersonaHub $0.81 \pm 0.12$
/ $0.69 \pm 0.09$ ($p = 0.02$ for trait); Anthology $1.06 \pm 0.09$ / $0.85 \pm 0.14$; Twin-2K-500
$0.95 \pm 0.05$ / $0.94 \pm 0.05$ (all values in Fig.~\ref{fig:sf-gating-conditions}).

\subsection{Network statistics}

Per-world contact networks are built based on the 56 h contact relation, i.e., two agent nodes are connected if any agent messaged, referred to, or read another's content in an hourly window. That means contact is undirected and includes passive exposure, reflecting real-life networks of mutual information exchange on social platforms. Each network is summarized by the statistics below, reported as mean $\pm 1$ SD over the ten worlds, following the conventions of \citet{papachristou2025network, sodano2026fragility, demarzo2023emergence}. Density is the fraction of agent pairs that are in contact. Transitivity $C$ is the fraction of connected agent node triples that close into triangles, i.e., how often two contacts of the same agent are also in contact with each other. Giant component is the share of agents in the largest connected subgraph. Path length $L$ is the mean shortest-path distance between agents in the giant component. $C/C_{\mathrm{rand}}$ and $L/L_{\mathrm{rand}}$ divide subgraphs by their values in Erdős-Rényi random graphs with the same number of nodes and edges (100 draws per world). Their ratio is the small-world coefficient $\sigma = (C/C_{\mathrm{rand}})/(L/L_{\mathrm{rand}})$, which exceeds 1 when a network is more clustered than chance. Modularity $Q$ measures how cleanly the network splits into subgraphs with more internal compared to external links. Degree assortativity is the correlation between the degrees of connected agent nodes. Direct-message reciprocity is the fraction of directed message ties that are returned by the recipient, and is computed on the directed message graph rather than the undirected contact graph.

\subsection{Human example study comparison}

We re-analyzed data originally published by \citet{isella2011crowd}, in which 113 conference attendees' RFID badges were tracked to register participant contact when two badges were within 1--1.5~m and facing each other
for a 20~s interval. The public file holds 20818 anonymous contact rows
over 59 clock hours, 43 of which contain any contact. The published data were analyzed using the same code that computes
our agent-world network parameters. First, time was cut into hourly windows, a pair
was considered in contact in a window if at least one 20~s interval fell into it (the
analogue of one message or one read making an hour a contact window for
agents), hours were re-indexed over the hours in which any contact occurred
so that the nights the attendees spend away from the venue do not count as
inactivity, and a pair was considered a sustained dyad if it shared $\geq 7$ such windows.
Using both the contact graph ($\geq 1$ window) and the sustained graph ($\geq 7$), we
computed contact density, transitivity,
average shortest path length on the giant component, the giant-component
fraction, $L/L_{\mathrm{rand}}$ and small-world $\sigma$ against
Erd\H{o}s--R\'enyi references, Louvain modularity $Q$, sustained dyads per
person, and pair persistence $P(\text{contact at }h{+}1 \mid h)$. As DM
reciprocity and platform use have no analogue in
undirected proximity data, they were not compared. Because an agent world has 50
agents in one shared space, parameters are computed on 50-person groups
drawn from the attendee network (200 draws). The reported value
is a connected-random group, consisting of a random starting person node, followed by repeated node draws adhering to the current contact frontier with contact-window-proportional probability.
Note that we consider physical face-to-face contact to be comparable to social platform contact for these analyses and comparisons.
Raw contact data obtained were released by the SocioPatterns collaboration \citep{sociopatterns2009ht09} under
CC~BY-NC-SA license.

\subsection{Action-space Markov chain}

Every tool call in every agent turn was mapped from the checkpoint archives to
a platform-explicit action state and stepped action-to-action, pooled over the
ten worlds and re-estimated per world to estimate standard deviations. We required $\geq 20$ visits of the source state for inclusion.
For visibility, ring transitions and poll returns below
7.5\% and exits to stop below 5\% are omitted, and the harness heartbeat
returning stop to wake is left implicit, except for memory-write states (Fig.~\ref{fig:sf-action}a). Similarly, we combined flows of the wake step and subsequent polls into one merged poll state, with waking straight to an immediate stop counted
as nothing new, and transitions back into the reading state folded into
each state's return to poll. Every state retains the inflow that feeds it the most visits, drawn dotted when below threshold. Compare Figs.~\ref{fig:sf-action},~\ref{fig:sf-action-gemini},~\ref{fig:sf-action-llama}, and~\ref{fig:sf-action-glm} for action-space Markov chains of Nemotron persona agents using Qwen3.5-122B-A10B, Gemini 3.8 Flash, Llama 4 Scout, and GLM-4.5-Air, respectively.

\subsection{Personality-distance analysis}

Each pair's 10-world formation count ($k$ of 10) is regressed on the euclidean distance between battery profiles
($z$-scored metrics, recomputed per world from that world's own sweeps), 
LLM-judged personality profiles, and cosine similarity of the persona system prompt (LLaMA-3.1-8B layer-16 mean-pooled embeddings). To analyze whether future sustained dyad formation can be predicted from any of the above relative personality space metrics, we utilized 15 quantile-spaced predictor bins with pooled Wilson 95\% CIs and
a predictor-density strip, overlaid with a weight-10 IRLS logistic fit and a
$\pm 2$ SEM band over the ten per-world fits. Slopes for personality batteries
are $-0.003 \pm 0.072$ log-odds per SD and
$-0.021 \pm 0.065$ (both null, rank tests $p > 0.75$), judged-personality distance is
$-0.140 \pm 0.049$ per SD, and persona-embedding
similarity is $\tau_b = +0.032 \pm 0.010$ ($p = 0.011$). Only the
persona-similarity rank test survives a Bonferroni threshold to correct for multiple comparisons.

\subsection{Personality and average expression}

Each agent's average expression is its speech-channel probe mean per world,
averaged over the ten repetition worlds. Personality coordinates are the top five PCs of
the bin-0 personality battery space, explaining 13.1 / 8.2 / 5.9 / 4.5 / 3.9\% of variance, respectively. Each of the 252
axes is tested by an omnibus OLS/ANOVA $F$-test of the five-PC regression
(df $= 5, 44$) against a Bonferroni threshold of
$\alpha = 0.05/252$. We also compared against a shuffled-null condition, in which we shuffled agent-labels. 201 of the 252 axes exceed the Bonferroni threshold, with mean $R^2 = 0.58$ compared to $R^2 = 0.10$ for the shuffled-null.

\subsection{Formation latency}

Although dyad overall formation
likelihood is predictable, agents' first-contact expression alignment does
not predict how quickly the dyad forms (per-world Spearman over the ten worlds was
$\rho = -0.053 \pm 0.049$, $p = 0.30$).

\subsection{Held-out prediction of sustained dyads}

To test whether first-contact expression improves out-of-sample prediction of sustained dyads over
simpler baselines, every first-contact pair of the ten reference worlds ($n = 473$--$695$ per world) was
labelled by whether it later became a sustained dyad, and logistic models were trained on nine worlds
and scored on the tenth (AUC on the held-out world, mean $\pm$ SD over the ten folds). Feature sets:
hour of first contact; persona-embedding cosine similarity; the two agents' pre-contact activity
(messages and distinct partners, sum and difference); first-contact alignment and extremity (emotion and
trait cosine alignment and joint deviation, Fig.~\ref{fig:probes}c); and same-window interaction volume
and reciprocity. Held-out AUCs: hour $0.54 \pm 0.05$; persona similarity $0.55 \pm 0.05$; pre-contact
activity $0.55 \pm 0.07$; activity + persona $0.56 \pm 0.07$; alignment + extremity $0.65 \pm 0.05$;
activity + persona + alignment $0.66 \pm 0.05$; volume + reciprocity + activity $0.75 \pm 0.05$; all
non-expression features $0.75 \pm 0.05$; all features $0.76 \pm 0.04$. Alignment + extremity improves
on hour, persona similarity, and pre-contact activity by $+0.10$ each (paired over worlds, $p = 0.0002$,
$0.007$, and $0.014$; eight to ten of ten worlds improved) and by $+0.10$ over activity + persona
combined ($p = 0.002$), but adds $+0.001$ on top of same-window volume and reciprocity, whose permutation
importance ($-0.11$ AUC when shuffled) dominates every other feature ($\leq 0.02$). Expression at first
contact therefore predicts beyond who the agents are and how active they were before meeting, but not
beyond how much they talk once they have met.

\end{document}